%% file: manuscript.tex
\documentclass[final,3p,times,twocolumn]{elsarticle}

\newcommand{\PL}{P{\tiny L} }
\newcommand{\PR}{P{\tiny R} }

\newcommand{\PLc}{P{\tiny L}}
\newcommand{\PRc}{P{\tiny R}}

\newcommand{\OM}{O{\tiny M} }
\newcommand{\OMc}{O{\tiny M}}

\newcommand{\OL}{O{\tiny L} }
\newcommand{\OLc}{O{\tiny L}}

\newcommand{\OD}{O{\tiny D} }
\newcommand{\ODc}{O{\tiny D}}

\newcommand{\OR}{O{\tiny R} }
\newcommand{\ORc}{O{\tiny R}}

\newcommand{\CI}{{\rm CI}}
\newcommand{\MOR}{{\rm MOR}}

\usepackage{natbib}

\usepackage{graphicx}

\usepackage{amssymb}

\journal{Journal of Molecular Biology}

\begin{document}

\begin{frontmatter}

\title{Modeling of the genetic switch of bacteriophage TP901-1: 
A heteromer of CI and MOR ensures robust bistability}

\author[Kyushu]{Hiizu Nakanishi}
\author[NBI]{Margit Pedersen}
\author[NBI]{Anne K. Alsing}
\author[NBI]{and Kim Sneppen}

\address[Kyushu]{Department of Physics, Kyushu University 33, Fukuoka
 812-8581, Japan} 
\address[NBI]{Center for Models of Life, Niels Bohr
 Institute, University of Copenhagen, 
Blegdamsvej 17, DK-2100 Copenhagen, Denmark}

\begin{abstract} 

The lytic-lysogenic switch of the temperate lactococcal phage TP901-1 is
fundamentally different from that of phage lambda.  In phage TP901-1,
the lytic promoter \PL is repressed by CI whereas repression of the
lysogenic promoter \PR requires the presence of both of the
antagonistic regulator proteins, MOR and CI.
We model the central part of the switch and compare the two cases for
\PR repression: the one where the two regulators interact only on the
DNA, and the other where the two regulators form a heteromer complex in
the cytoplasm prior to DNA binding.
The models are analyzed for bistability, and the predicted promoter
repression folds are compared to experimental data.
We conclude that the experimental data are best reproduced the latter
case, where a heteromer complex forms in solution.  We further find that
CI sequestration by the formation of MOR:CI complexes in cytoplasm makes
the genetic switch robust.

\end{abstract}


\end{frontmatter}

\section{Introduction}\label{intro}

Phenotypic variability under homogeneous conditions can readily be
obtained by interlinking multiple gene regulatory pathways. Several
well-characterized examples of phenotypic variations are known to be
important for different developmental process of bacteria, such as the
presence of persister cells in {\it Staphylococcus aureus} and {\it
E. coli}, development of natural competence and sporulation in {\it
Bacillus subtilis}, and the choice between lytic or lysogenic growth of
temperate bacteriophages \cite{dubnau2006, veening2008}. Two
distinguishable phenotypes may originate from a bistable system, i.e. a
system that can toggle between two alternative stable steady-states
\cite{ferrell2002}. Infection of bacteria by temperate bacteriophages
provides a classical example of the possibility to choose between two
alternative modes of development.

The bacteriophage lambda infecting {\it Escherichia coli} has been
subjected to decades of intensive study, making the lytic-lysogenic
switch one of the best understood gene regulatory
systems\cite{Eisen1970,Svenningsen2005,Ptashne2004}. The bistability of
the lambda switch is obtained from a double negative feedback mechanism,
where two repressor proteins directly repress transcription of the other
repressor gene. This system has a stable state with one promoter on and
the other off, and vice versa for the other stable state. Once either
state has been established, it would persist indefinitely or until some
trigger stimulus forces the system to switch to the other
state.

\begin{figure}
\includegraphics[width=0.5\textwidth]{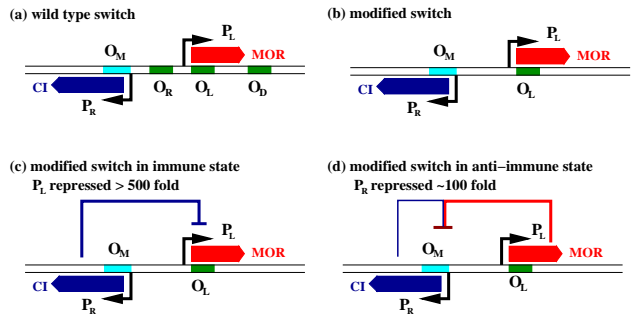} 
\caption{The regulatory circuit of the genetic switch from phage TP901-1
may be described as a double negative regulatory network.  (a) Wild type
genetic switch with \OLc, \ORc, and \OD sites.  The relative positions
of the {\it cI} and {\it mor} genes and the divergently oriented
promoters \PR and \PL are also shown.  The cyan box represents a
putative \OM site.  (b) Modified switch investigated in this paper.
Only one CI operator site, \OLc, is present (dark green box).  (c)
Immune state of the modified switch. CI represses transcription from \PL
approximately 1000-fold by binding to the \OL operator site with
transcription from \PR being allowed.  (d) Anti-immune state of the
modified switch.  Repression of \PR depends on both CI and MOR, which
repress transcription from \PR approximately 100-fold with transcription
from \PL being allowed.  }
\label{fig:circuit}
\end{figure}

The genetic switch of the temperate lactococcal bacteriophage TP901-1
infecting {\it Lactococcus lactis} subsp. {\it cremoris} provides a
regulatory system diverse from the lambda genetic switch.  A previous
study has demonstrated that a DNA fragment obtained from the temperate
lactococcal phage TP901-1 shows bistability when introduced
into {\it Lactococcus lactis}. The cloned DNA fragment contains the two
divergently oriented promoters, \PR and \PLc, and the two promoter
proximal genes {\it cI} and {\it mor}
\cite{pedersen2008b}(Fig.\ref{fig:circuit}a). A knockout mutation in the
{\it mor} gene showed that CI ensures tight repression of the \PL
promoter and partially repression of the \PR promoter whereas a knockout
mutation in the {\it cI} gene results in open states of both \PR and
\PLc, showing that MOR by itself does not exhibit repression of either
promoter\cite{madsen1999, pedersen2008a}.
Two types of repression has been shown: i) {\em MOR-independent
repression}, which is responsible for repression of \PLc. The \PL
promoter is repressed by cooperative binding of CI to the three operator
sites \ORc, \OL and \ODc, by the formation of a CI-DNA loop
structure. ii) {\em MOR-dependent repression}, which is responsible for
repression of \PR and only occurs in the presence of both MOR and
CI. This repression is suggested to occur through MOR and CI binding at
a putative \OM operator site \cite{pedersen2008a}. Hence, the
bistability of the genetic switch from phage TP901-1 may be described as
a mixed feedback loop, 
where both of the antagonistic repressor proteins are involved in \PR
repression.
It is still not clear how MOR and CI collectively repress transcription
from \PRc, and so far there is no direct experimental evidence for
interaction between CI and MOR.

In order to understand the mechanism of switching in phage TP901-1, we
here study a modified version of the cloned wild-type TP901-1
switch(Fig.\ref{fig:circuit}b).  This construct contains only one of the
three CI operator sites, \OLc, which gives tight repression of \PL thus
still sustains the bistable behavior of the construct.  In the immune
state, the \PL promoter is repressed approximately 1,000-fold, but high
expression from \PR is allowed due to the absence of \ORc, that
autoregulates transcription from \PR in the case of the wild-type switch
\cite{pedersen2008a} (Fig.\ref{fig:circuit}c). In the anti-immune state,
\PR is repressed approximately 100-fold but high expression from \PL is
allowed\cite{pedersen2008a} (Fig.\ref{fig:circuit}d).

Generally speaking, it is not easy to predict the behavior of a bistable
switch without quantitative analysis because a bistable switch is a
dynamical and highly nonlinear system.  In the present case of
TP901-1, even the modified version of the switch (Fig.\ref{fig:circuit}b) could
involve a number of mechanisms, such as cooperativity binding via
homo/hetero-dimerization, sequestration via heterodimerization,
intertwined loops of negative and positive feed back via protein
interactions, etc.  In such a situation, the only way to obtain any reliable
results is to perform quantitative analysis on specific models.  By
confronting numerical results with experimental data, we can restrict
possible mechanisms with plausible parameters

\begin{figure}[t]
\centerline{
\includegraphics[width=0.45\textwidth]{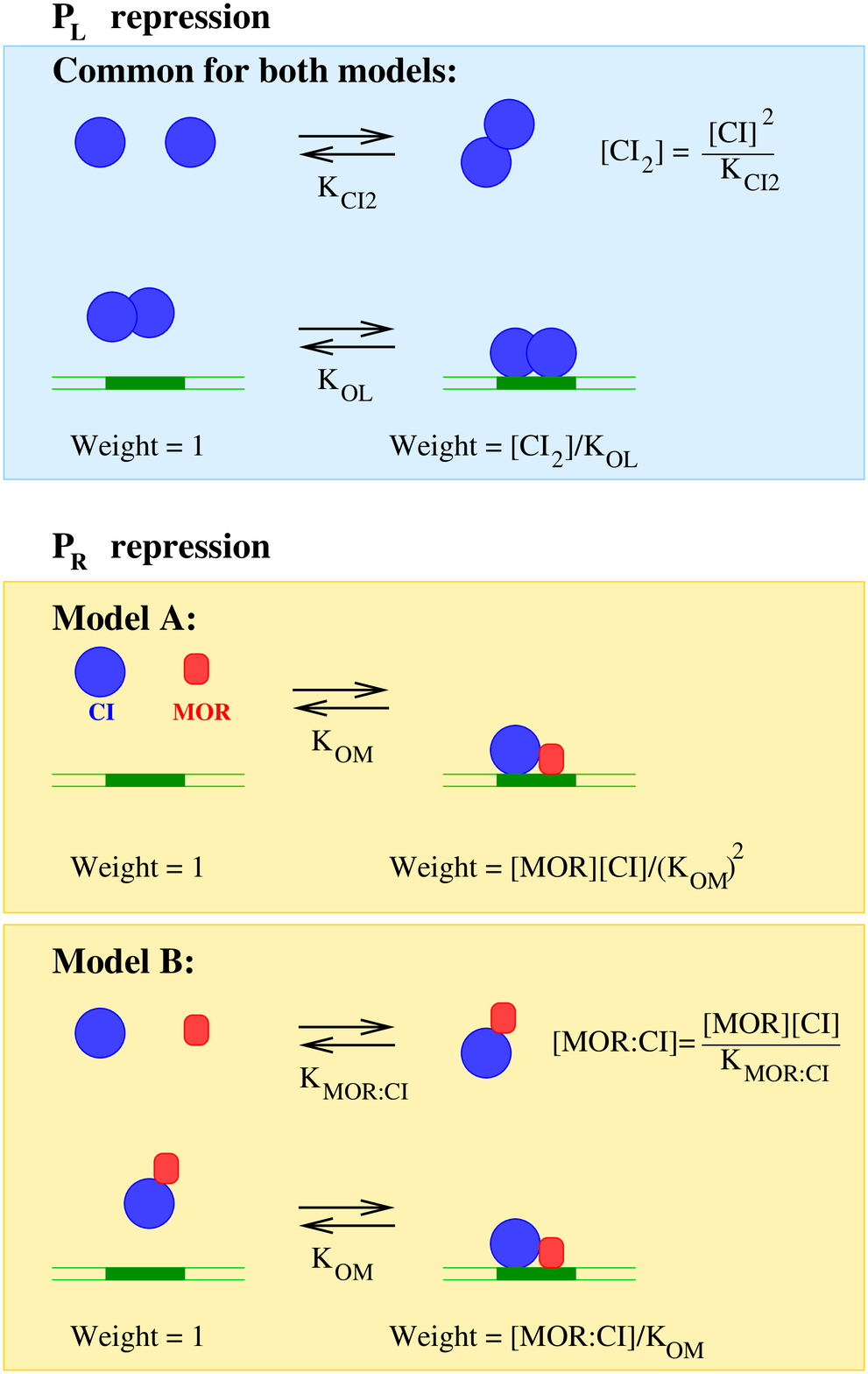}}
\caption{Repression of \PL though binding of $\CI_2$ at \OL is employed
in both models (blue box). Repression of \PR is supposed to occur by
direct binding of CI and MOR to \OM (Model A) or by formation of a
MOR:CI complex prior to binding to \OM (Model B). The statistical weight
is given for the different states of the operator sites. In Model A, it
is assumed that binding of CI alone to \OM is so week that the operator
is only occupied when binding is stabilized by MOR, or vice versus if
MOR binds first.}  \label{fig:models}
\end{figure}

In this paper, we construct mathematical models for this modified
bistable system based upon statistical mechanics, examine their behavior
in the steady states numerically, and compare the obtained repression
folds with experimental observations\cite{pedersen2008a}.  We assume
that \PL is repressed by the CI dimer binding to the operator \OL since
the operator has two inverted repeated sequences \cite{johansen2003}.
On the other hand, the MOR-dependent repression of \PR is assumed to be
brought about by the MOR:CI:DNA complex formation at the putative \OM
operator.
Since the amino sequence of MOR shows high similarity to the DNA-binding
Helix-turn-Helix domain of the repressor protein encoded by {\it
Escherichia coli} phage 434 \cite{Shimon1993}
MOR is likely to be a DNA binding protein, but
the fact that MOR alone does not repress transcription from either \PL
or \PR suggests that the DNA-binding affinity of MOR is negligible.  CI
alone does not repress \PR in this system due to absence of \OR
operator.  Based upon these observations, we test two different
scenarios for \PR repression(Fig.\ref{fig:models}): i) repression
through direct binding of CI and MOR to \OM (Model A), and ii)
repression through binding of MOR:CI complex to \OM with the complex
being formed in the cytoplasm prior to the binding to DNA (Model B).

The MOR:CI:DNA complex formation in Model A may be regarded as an
extreme case of Model B where the MOR:CI complex formation in cytoplasm
is so weak that the complex is stabilized only when it binds to DNA,
thus there is no substantial presence of MOR:CI complex in the
cytoplasm.  However, distinguishing Model A from Model B helps us to
recognize two distinct aspects in the \PR repression by
the MOR:CI:DNA complex, i.e., co-operativity and sequestration;
The latter has been studied in silico as a possible mechanism for
a genetic switch\cite{Francois2004,Francois2005} and demonstrated to
provide strong nonlinearity\cite{Buchler2008,Buchler2009}.  In fact,
our Model B is a reminiscent of one of the bistable switches obtained by
the simulated evolution (Fig.3A in \cite{Francois2004}). 
We will demonstrate that CI sequestration by the MOR:CI formation in
cytoplasm can make a robust bistable system and is actually a plausible
switching mechanism for TP901-1.

\section*{Theory}

The regulatory circuit in the present system consists of the two
promoters \PL and \PRc, which produce MOR and CI,
respectively(Fig.\ref{fig:circuit}b).  The promoter \PL is repressed by
CI binding at \OLc, thus the \PL activity is given by a function of the
CI concentration as
\begin{equation}
pL([\CI]) = pL_0 \cdot f_{\rm OL}([\CI]),
\label{pL-0}
\end{equation}
where $pL_0$ is the bare activity of the promoter \PLc. The function
$f_{\rm OL}([\CI])$ represents the repression factor. In the absence of CI,
there is no repression: $f_{\rm OL}(0)=1$.  The \PR activity, on the other
hand, depends on the concentrations of both CI and MOR.  Accordingly,
the \PR activity can be written as
\begin{equation}
pR([\MOR], [\CI]) = pR_0\cdot f_{\rm OM}([\MOR], [\CI])
\label{pR-0}
\end{equation}
with $pR_0$ being the bare activity.  The function $f_{\rm
OM}([\MOR],[\CI])$ is the repression factor due to the binding of MOR
and CI at \OMc, and satisfies $f_{\rm OM}(0,[\CI])=f_{\rm
OM}([\MOR],0)=1$.

The promoter \PL produces MOR, and \PR produces CI, thus in the modeled
feedback system, the total concentration for each protein, $[\MOR]_{\rm
total}$ and $[\CI]_{\rm total}$, is governed by the dynamics equations,
\begin{equation}
 {d\over dt}[\MOR]_{\rm total}  = 
{1\over \tau_M}\Bigl( pL([\CI]) - [\MOR]_{\rm total} \Bigr),
\label{t-dep-MOR}
\end{equation}\begin{equation}
 {d\over dt}[\CI]_{\rm total}  =  
{1\over \tau_C}\Bigl( pR([\MOR], [\CI]) - [\CI]_{\rm total} \Bigr),
\label{t-dep-CI}
\end{equation}
where $\tau_M$ and $\tau_C$ are the degradation times for MOR and CI,
respectively.  To simplify the notation, we have rescaled the promoter
activities, Eqs.(\ref{pL-0}) and (\ref{pR-0}), by the degradation times,
i.e.  the promoter activities are now measured in terms of the steady
state protein concentrations.

In steady states, the production and the degradation of each protein
should balance, therefore, the promoter activities and the concentrations
of the expressed proteins in the cytoplasm should satisfy {\em the steady
state condition},
\begin{eqnarray}
pL([\CI]) & = & [\MOR]_{\rm total},
\label{SCE-pL}
\\
pR([\MOR], [\CI]) & = & [\CI]_{\rm total}.
\label{SCE-pR}
\end{eqnarray}

Not all steady states are stable against small perturbations.  A steady
state is unstable if a perturbation drives the system out of the state;
The stability should be determined by the dynamics equations,
Eqs.(\ref{t-dep-MOR}) and (\ref{t-dep-CI}) (See supplementary material).
If there are two stable steady states, the system shows bistability.
\\

We assume that the repression factors, $f_{\rm OL}$ and $f_{\rm OM}$ in
Eqs.(\ref{pL-0}) and (\ref{pR-0}), are given by the statistical weights
at equilibrium that the corresponding operators are not bound by the
regulators. This approximation holds when the time that RNA
polymerase (RNAP) needs to start elongation after binding to DNA is much
shorter than the time scales of binding/unbinding of RNAP and repression
factors to the promoter/operator sites \cite{nakanishi2008}.  
The equilibrium statistical weights depend upon the repressor
concentrations, and their dependence is characterized by the Hill
coefficient and the affinities of the repressors to the operator sites
\cite{ackers1982,sneppen2005,dodd2007}.
\\

For {\em the MOR-independent repression of \PLc}, we suppose that \PL is
repressed by CI dimer binding at \OLc, and that the dimers are formed in
the cytoplasm before binding. Thus, within the above approximation for
the repression factor, the \PL activity is given by
\begin{equation}
pL({\rm [\CI]})  =  pL_0 \cdot  {1\over 1 + [\CI_2]/K_{\rm OL}},
\label{pL-CI_2}
\end{equation}
where the affinity $K_{\rm OL}$ represents the CI$_2$ concentration at
which \OL is occupied for 50\% of the time.  Since the dimer concentration
[CI$_2$] is related to the monomer concentration [CI] as
\begin{equation}
[\CI_2]   =  {[\CI]^2\over K_{\rm \CI_2}}
\label{CI_2}
\end{equation}
with the dissociation constant $K_{\rm \CI_2}$,
Eq.(\ref{pL-CI_2}) may be written as
\begin{equation}
pL({\rm [\CI]})  =  pL_0 \cdot  {1\over 1 + [\CI]^2/\tilde K_{\rm OL}^2}
\label{pL-CI}
\end{equation}
with the effective affinity 
\begin{equation}
\tilde K_{\rm OL} \equiv \sqrt{K_{\rm \CI_2} K_{\rm OL}}
\label{tilde_K_OL}
\end{equation}
for CI concentration.  In Fig.\ref{Model-A}, the
activity of \PL as a function of [CI] is plotted by a green line.
\\

As for {\em the MOR-dependent repression of \PRc},
we will examine two models.
In Model A, monomers of MOR and CI may bind cooperatively at \OMc, but
we do not assume any MOR:CI complexes formed in cytoplasm before binding
to DNA.  In Model B, on the other hand, CI and MOR may associate in
cytoplasm before they bind at \OMc.  For both models, \PR is repressed
by the formation of the MOR:CI:DNA complex at \OMc.  The important point
in Model B is that {\em the formation of MOR:CI heteromers competes with
CI dimer formation by sequestering CI monomers }.

\subsection*{Model A}
We first consider a MOR:CI:DNA complex containing one MOR and one CI
protein as illustrated in Fig.\ref{fig:models}.  Then, we can 
approximate the total concentration of MOR unit by the MOR monomer
concentration,
\begin{equation}
[\MOR]_{\rm total} = [\MOR].
\label{MOR_total-MOR-A}
\end{equation}
The activity of the \PR promoter is repressed from the bare activity
$pR_0$ by the statistical weight that the operator \OM is not occupied by
MOR and CI,
\begin{equation}
pR([\MOR],[\CI])  =  pR_0 \,{1\over 1+ [\MOR][\CI]/(K_{\rm OM})^2}.
\label{pR-A}
\end{equation}
The affinity $K_{\rm OM}$ is the concentration $\sqrt{[\MOR]\cdot [\CI]}$
where \OM is occupied by MOR and CI for 50\% of the time. 

The steady state is determined from the steady state condition
Eqs.(\ref{SCE-pL}) and (\ref{SCE-pR}) by eliminating the MOR
concentrations.  With the help of Eq.(\ref{MOR_total-MOR-A}), we obtain
\begin{eqnarray}
pR\Bigl( pL([\CI]),[\CI] \Bigr) & = & [\CI]_{\rm total},
\label{pR-CI_total-A}
\end{eqnarray}
which represents the balance between the production and the degradation
of CI.  This can be solved graphically by plotting the both sides as a
function of [CI],
\begin{eqnarray}
\lefteqn{
pR\Bigl( pL([\CI]),[\CI] \Bigr)  }\hfill \nonumber \\
&= &
pR_0\left[
1+  { pL_0\, \tilde K_{\rm OL}/(K_{\rm OM})^2
\over ([\CI]/\tilde K_{\rm OL}) + (\tilde K_{\rm OL}/[\CI]) }
\right]^{-1},
\label{pR-pL-CI-A}
\end{eqnarray}\begin{equation}
{[\CI]_{\rm total}} =  [\CI] +  2 {[\CI]^2\over K_{\CI_2}}.
\label{CI_total-CI-A}
\end{equation}
Eq.(\ref{pR-pL-CI-A}) represents the \PR activity in the system where
MOR is provided by \PL but [CI] is controlled externally.  Note that the
relative strength of the bare promoters, $pL_0$ and $pR_0$, does not
affect the system behaviors, such as bistability or repression folds,
because there is no direct interaction between MOR and CI in this
model.

\subsection*{Model B}

In this model, a $\MOR\;\CI$ heterodimer is formed in solution before it
binds to the putative \OM site to repress \PR (Fig.\ref{fig:models}).
The activity of the \PL promoter is again given by Eq.(\ref{pL-CI}) but
the \PR activity is
\begin{equation}
pR([\MOR],[\CI])  =  pR_0 \,{1\over 1+ [\MOR\;\CI]/K_{\rm OM} },
\label{pR-B}
\end{equation}
where $K_{\rm OM}$ now represents the concentration of the $\MOR\;\CI$
heterodimer at which \OM is occupied for 50\% of the time.  The
concentration of the $\MOR\;\CI$ heterodimer is given as
\begin{equation}
{[\MOR\;\CI]}  =  {[\MOR]\cdot [\CI]\over K_{\MOR\;\CI}}
\label{MOR-CI-B}
\end{equation}
with the dissociation constant $K_{\MOR\;\CI}$ for the heterodimer.

The formation of the heterodimers couples the monomer concentrations of CI
and MOR through
\begin{eqnarray}
{[\CI]_{\rm total}} & = & [\CI] + 2[\CI_2] + [\MOR\;\CI],
\label{CI_total-B}
\\
{[\MOR]_{\rm total}} & = &  [\MOR] + [\MOR\;\CI],
\label{MOR_total-B}
\end{eqnarray}
which leads to the competition between the $\CI_2$ formation and the
$\MOR\;\CI$ formation.
Note that $[\CI_2]$ is still given by Eq.(\ref{CI_2}).

The steady state is determined as in the case of Model A;  We consider
the \PR activity as a function of [CI] when MOR is provided by \PLc.
From Eqs.(\ref{MOR-CI-B}) and (\ref{MOR_total-B}), [MOR] is expressed in
terms of [CI] and $[\MOR]_{\rm total}$,
\begin{equation}
[\MOR] = {[\MOR]_{\rm total}\over 1 + [\CI]/K_{\MOR:\CI}},
\label{MOR-B}
\end{equation}
and then, in the steady state where Eq.(\ref{SCE-pL}) holds, $[\MOR]_{\rm
total}$ is given by the \PL activity with Eq.(\ref{pL-CI_2}).
Then the steady state condition Eq.(\ref{SCE-pR}) for \PR becomes
\begin{equation}
pR\left({pL([\CI])\over 1 + [\CI]/K_{\MOR:\CI}},\, [\CI]\right)
   =  [\CI]_{\rm total},
\label{SCE-B}
\end{equation}
which can be solved graphically with the explicit forms for the both sides:
\begin{eqnarray}
\lefteqn{ \hspace{-2em}
pR\left({pL([\CI])\over 1 + [\CI]/K_{\MOR:\CI}},\, [\CI]\right)
= pR_0\Biggl[1+ } \nonumber\\
& & \hspace{-2em}
 {pL_0\, \tilde K_{\rm OL}/(\tilde K_{\rm OM})^2
\over ([\CI]/\tilde K_{\rm OL}) + (\tilde K_{\rm OL}/[\CI])}\cdot
{1\over 1 + [\CI]/K_{\MOR:\CI}}
\Biggr]^{-1},
\label{pR-pL-CI-B}
\end{eqnarray}
\begin{equation} \hspace{-2em}
{[\CI]_{\rm total}}  = 
[\CI] + 2 {[\CI]^2\over K_{\CI_2}} 
 +pL([\CI])\, {[\CI]/K_{\MOR:\CI} \over 1+ [\CI]/K_{\MOR:\CI}},
\label{CI_total-CI-B}
\end{equation}
where the effective affinities are
\begin{equation} \hspace{-2em}
\tilde K_{\rm OM} \equiv \sqrt{K_{\MOR:\CI}\cdot K_{\rm OM}}
\;\mbox{and}\;
\tilde K_{\rm OL} \equiv  \sqrt{K_{\CI_2} K_{\rm OL}}.
\label{tilde_K_OM}
\end{equation} 
Eq.(\ref{SCE-B}) represents the balance between the production and
degradation of CI in the system where MOR is provided by \PLc.
Note that Model B reduces to Model A in the limit of large
$K_{\MOR:\CI}$ with $\tilde K_{\rm OM}$ being kept constant
as has been discussed at the end of Introduction.

\section{Results}

We numerically examine the steady states for the two versions of the
models we have constructed (Fig.\ref{fig:models}).

\begin{figure}
\centerline{
\includegraphics[width=0.5\textwidth]{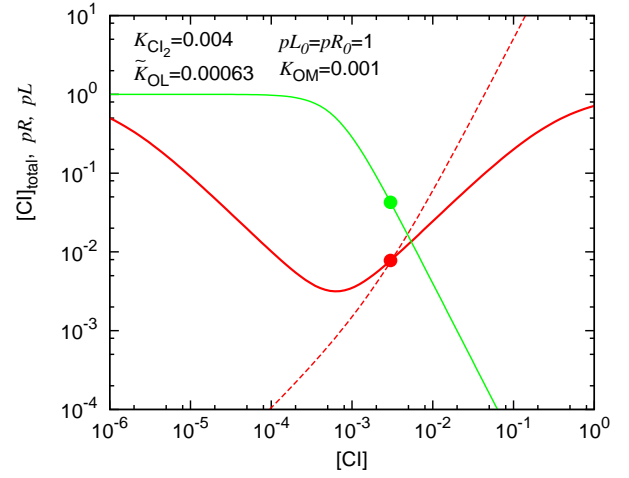} }
\caption{Model A. Promoter activities, $pL$ and $pR$, and $[\CI]_{\rm
total}$ as a function of free [CI]. Curves represent \PL activity of
Eq.(\ref{pL-CI}) (green line), \PR activity of Eq.(\ref{pR-pL-CI-A})
(solid red line), and $[\CI]_{\rm total}$ of
Eq.(\ref{CI_total-CI-A})(dashed red line).  The filled red circle at
the intersection between the [CI]$_{\rm total}$ and \PR curve
represents the steady state. }  
\label{Model-A} 
\end{figure}

\subsection*{Model A}

In our first model, we study the possibility for bistability in the
system where \PR is repressed by binding a MOR monomer and a CI monomer
to \OM without direct interaction between MOR and CI in the cytoplasm.
The binding affinities for each of the proteins alone at \OM should be
negligible because the \PR repression requires both of the proteins.
Hence, the affinity $K_{\rm OM}$ in Eq.(\ref{pR-A}) 
may be considered as the effective binding affinity $\tilde K_{\rm OM}$
for MOR and CI with very weak $\MOR\;\CI$ formation, or as
the resulting binding affinities from CI:\OMc, MOR:\OMc, and the
interaction between the bound proteins.

Figure \ref{Model-A} shows a typical example of the \PL activity of
Eq.(\ref{pL-CI}) (green line), the \PR activity of Eq.(\ref{pR-pL-CI-A})
(solid red line), and the total concentration of CI, or degradation
rate, of Eq.(\ref{CI_total-CI-A}) (dashed red line) as a function of
free [CI] in the logarithmic scale.  One can see that \PL (green line)
is fully active and produces a lot of MOR at low [CI], whereas its
activity is monotonically decreasing with increasing [CI], due to \PL
repression by $\CI_2$ binding at \OLc; \PL is virtually shut down beyond
$[\CI]\approx \tilde K_{\rm OL} = \sqrt{K_{CI_2} \cdot K_{\rm OL}} = 6
\cdot 10^{-4}$.

On the other hand, the \PR activity (solid red line) shows a more
complicated behavior, i.e., \PR is open both at very low and high [CI]
concentrations but repressed at the intermediate concentration.  One can
understand this behavior by noting that the solid red line,
Eq.(\ref{pR-pL-CI-A}), represents the \PR activity in the system where
MOR is expressed from \PL under CI control; At low [CI], there is plenty
of MOR due to high \PL activity, while at high [CI], no MOR is
present. The \PR activity is repressed only at intermediate [CI] because
both MOR and CI are necessary for its repression by means of the
MOR:CI:DNA complex formation.

In the steady state, the production and the degradation of each protein
should balance.  Since CI is expressed from \PR and the degradation of
CI is assumed to be proportional to the total concentration of CI, the
steady states are identified as $pR(pL({\rm [CI]}), {\rm [CI]}) = {\rm
[CI]}_{\rm{total}}$, namely, Eq.(\ref{pR-CI_total-A}).  We thus find the
steady states at the intersection points between the $pR$ and ${\rm
[CI]}_{\rm{total}}$ curves represented by Eqs.(\ref{pR-pL-CI-A}) (solid
red line) and (\ref{CI_total-CI-A}) (red dashed line), respectively.

In Fig.\ref{Model-A}, only one intersection point is observed between
the two curves, showing that there is only one steady state solution.
This uniqueness of steady state holds true for any given value of the
parameters, because the activity of \PR given by Eq.(\ref{pR-pL-CI-A})
never increases faster than proportional to \CI, whereas
[CI]$_{\rm{total}}$ by Eq.(\ref{CI_total-CI-A}) always increases faster
or proportional to [CI]. Therefore, bistability is never realized in
Model A with the assumption of binding of one MOR and one CI for
repression of \PRc.

\subsection*{Variant of Model A}

\begin{figure}
\centerline{\includegraphics[width=0.5\textwidth]{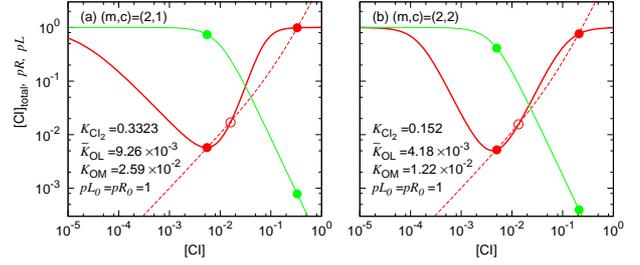}}
\caption{ Model A with 
(a) two MOR monomers and one CI monomer binding to \OMc,
$(m,c)=(2,1)$,  and
(b) two MOR monomers and two CI monomers, $(m,c)=(2,2)$. Promoter
activities, $pL$ and $pR$, and $[\CI]_{\rm total}$ as function of free
[CI].  Curves are \PL activity of Eq.(\ref{pL-CI}) (green line), \PR
activity of Eq.(\ref{pR-A'}) (solid red line), and $[\CI]_{\rm total}$
of Eq.(\ref{CI_total-CI-A})(dashed red line).  The filled red circles at
the intersections represent the stable steady state, and the green
circles indicate the \PL activity in the steady states.  The open red
circles mark the unstable steady states.  The repression folds $pR({\rm
open})/pR({\rm closed})$ and $pL({\rm open})/pL({\rm closed}) $ are
approximately 200 and 1000 respectively.
} 
\label{Model-A'}
\end{figure}

Bistability may be obtained in Model A if a larger number of proteins
are allowed to form a complex structure at \OMc.  Suppose $m$ MOR
monomers and $c$ CI monomers bind at \OM to form the MOR$_m$:CI$_c$:DNA
complex that represses transcription from \PRc, then for the expression
for the \PR activity, Eq.(\ref{pR-A}) should be replaced by
\begin{eqnarray}
\lefteqn{ pR([\MOR],[\CI])  } \nonumber \\
&=&  pR_0 \,{1\over 1+ [\MOR]^m [\CI]^c/(K_{\rm OM})^{m+c}},
\label{pR-A'}
\end{eqnarray}
while all the other equations remain the same.  Using this for the left
hand side of the steady state equation (\ref{pR-CI_total-A}) with $[\MOR
] \propto 1/[\CI]^2$ from Eq.(\ref{pL-CI_2}) for large [CI], one can see
that the largest slope of the \PR activity as a function of [CI] is
$2m-c$ in the logarithmic scale.  Since the slope of the plot of
$[\CI]_{\rm total}$ given by Eq.(\ref{CI_total-CI-A}) is between 1 and
2, the steady state equation (\ref{pR-CI_total-A}) can have more
than two solutions with Eq.(\ref{pR-A'}) when $2m-c\ge 2$.  In the case
of $m=2$, we could obtain multiple solutions with $c=1$ or 2, i.e.,
two MOR monomers binding together with one or two CI monomers at \OMc.
Examples for Model A with $(m,c)=(2,1)$ and $(2,2)$ are shown in Fig.
\ref{Model-A'}. In each example, the intersections between the solid red
line (the \PR activity) and the dashed red line (the total CI
concentration) represent steady state solutions. One can see there are
three solutions for each case in Fig.\ref{Model-A'}.

Dynamical analysis shows that the steady state in the middle marked by an
open red circle is unstable against small fluctuations, and the
states at the ends marked by filled red circles are stable (See
supplementary material for detail).
Full analysis requires Eqs.(\ref{t-dep-MOR}) and (\ref{t-dep-CI}), but
the stability may be understood in the following way; For the steady
state in the middle, if the CI monomer concentration increases by
fluctuation from the steady value of [CI], the CI production from \PR
will increase more than the increase in degradation given by $[\CI]_{\rm
total}$, as is seen in Fig.\ref{Model-A'}, where the solid red line of
the \PR activity goes above the dashed red line of $[\CI]_{\rm total}$,
upon increasing [CI] from the middle steady state.  This means that such
a fluctuation causes further increases of [CI], consequently, the state
is driven out of the steady state.
On the other hand, the steady states at both ends represent
stable states. A fluctuation towards larger [CI] leads to
insufficient CI production in comparison with the CI degradation,
i.e. the solid red line goes under the dashed red line as [CI]
increases. This brings the system back to the original state,
therefore, they are stable.
Thus,
the system has two stable steady states, which leads to 
bistability.\footnote{
Note that this simplified analysis is based on the assumption that
Eq.(\ref{SCE-pL}) always holds, i.e. the response of \PL is much faster
than that of \PRc.  For general case, see supplementary material.}
%
\begin{figure}
\centerline{
\includegraphics[width=0.5\textwidth]{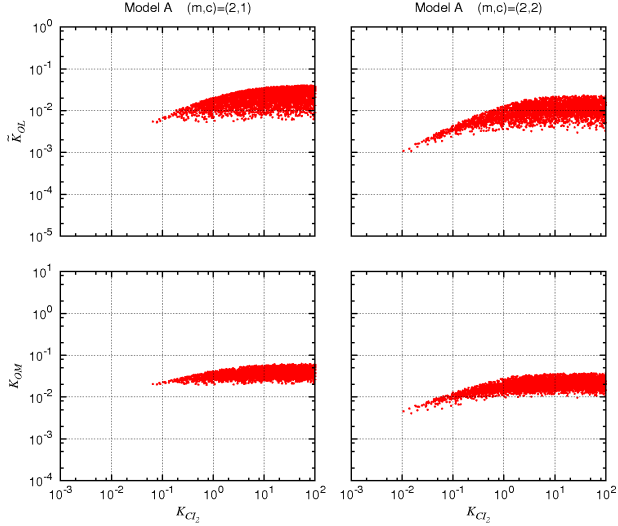}
}
\caption{Distributions of possible parameters for Model A.  
The plotted parameter sets are
those that give the repression folds $50<pR({\rm open})/pR({\rm
closed})<200$ and $pL({\rm open})/pL({\rm closed})>500$  out of
$10^6$ tested parameter sets, which are randomly picked  in the
logarithmic scale over the range of $\tilde K_{\rm OL}\in(10^{-5},10^{-2})$,
$K_{\rm OM}\in(10^{-4},10)$, and $K_{\CI_2}\in(10^{-3},10^2)$.  
}
\label{fig:param-A}
\end{figure}

The promoter activities in the two stable states can be
determined from the graphic representation in Fig.\ref{Model-A'}.
The \PR activity is read from the ordinates of the intersection (filled
red circles) and the \PL activity is read off from the corresponding
[CI] values of the intersection points (green circles), which allows us
to estimate the repression folds for \PL and \PR between the two stable
states.  The state at the right represents the immune state with open
\PR and repressed \PLc, while the one at the left represents the
anti-immune state with open \PL and repressed \PR.

The relative activities between the two states should be compared with
the promoter activities obtained from the {\it in vivo} measurements
\cite{pedersen2008a}; \PL is repressed approximately 1,000-fold in the
immune state and \PR approximately 100-fold in the anti-immune state.
To reproduce these repression folds in Model A with $(m,c)$ = (2,1) and
(2,2), we test the three parameters, $K_{\CI_2}$, $\tilde K_{\rm OL}$,
and $K_{\rm OM}$, representing the dimerization constant of CI, the
effective binding constant of CI-monomer at \OLc, and the binding
constant of the MOR:CI complex at \OMc, respectively, by setting 
{\em the criterion that the \PR repression fold should be in the range from
50 to 200 ( $50<pR({\rm open})/pR({\rm closed})<200$) and the \PL
repression fold should at least be 500 ($pL({\rm open})/pL({\rm
closed})>500$)}.  
Fig.\ref{fig:param-A} shows the distributions of accepted values of
parameters out of randomly chosen values in the logarithmic scale.
$\tilde K_{\rm OL}$ and $K_{\rm OM}$ are narrowly distributed while
$K_{\CI_2}$ are larger than $10^{-1}$ for $(m,c)=(2,1)$ and larger than
$10^{-2}$ for (2,2) in the unit of CI concentration at full activity of
\PRc.  One can also see that the accepted values for $K_{\CI_2}$ are
much larger than $\tilde K_{\rm OL}$.  This suggests that, in order for
Model A to work, CI must exist as a monomer and act by cooperative
binding to form $\CI_2$ at \OL when repressing \PLc.

Figure \ref{fig:ParamScan-Model-AB} shows the parameters that satisfy
the criterion only for the \PR repression fold versus resulting \PL
repression fold (left two columns for Model A and right two for Model
B).  The vertical green lines are drawn at the \PL repression fold 500,
thus only the plots on the right side of the lines should be accepted by
the repression fold criterion.
From the plots for $K_{\CI_2}$ for Model A, one can see that the
relatively high values for $K_{\CI_2}$ in this model comes from the
requirement for the large \PL repression fold.  This can be understood
as follows; In order to achieve large repression fold for \PLc, the
difference in [CI] for the two steady states should be large, which in
turn requires smaller slope in the $[\CI]_{\rm total}$ curve, namely
larger $K_{\CI_2}$ because the slope in $[\CI]_{\rm total}$ changes from
1 to 2 around $[\CI]\approx K_{{\CI_2}}$ (Fig.\ref{Model-A'}).


\subsection*{Model B}

Now, we consider the possibility that \PR is repressed by a MOR:CI
complex formed in cytoplasm before binding to DNA.
Examples for Model B are shown in
Fig.\ref{Model-B-m1c1}, where the \PL activity
of Eq.(\ref{pL-CI}) (green line), the \PR activity of
Eq.(\ref{pR-pL-CI-B}) (solid red line), and the total density of CI of
Eq.(\ref{CI_total-CI-B}) (dashed red line) as a function of [CI] in the
logarithmic scale; All of them have three steady states.  

Striking difference from the case of Model A is that the $[\CI]_{\rm
total}$ (dashed red line) can be non-monotonic.  Therefore, for a
given $[\CI]_{\rm total}$ within a certain range, there exist three
possible states with different [CI].
This suggests that the bistability could
be obtained for the system with \PL even if \PR were not regulated,
namely, even if \PR would produce CI at a fixed rate within the range.
This bistability is due to the MOR:CI heteromer formation in cytoplasm;
\PL produces MOR, which sequesters its own repressor, i.e. CI, by
forming MOR:CI.  For a given [\CI]$_{\rm total}$, the state at low [CI]
is the state where most of CI's are incorporated in MOR:CI heterodimers
due to the MOR produced by \PLc, while the state at high [CI] is
the state where most of CI is in the dimer with \PL being repressed.
The state in the middle is unstable.
Such bistability is, of course, not the bistability observed in the
experiments, but one can see that this feature of behavior in
$[\CI]_{\rm total}$ (dashed red line) makes it easier to have three
intersections with the \PR activity curve (solid red line) than in the
case of Model A.

According to the stability criterion we discussed, the steady state at
both ends are stable while the state in the middle is unstable even for
the system where both \PR and \PL are regulated.  Full analysis,
however, shows there are some cases where the states at both ends can be
unstable although the stability criterion is correct for most cases (See
supplementary material).  We analyse the bistability based upon the
stability criterion, ignoring the small possibility that the states at
both ends could be unstable.

\newpage

{\onecolumn
\begin{figure}[h]
\includegraphics[width=\textwidth]{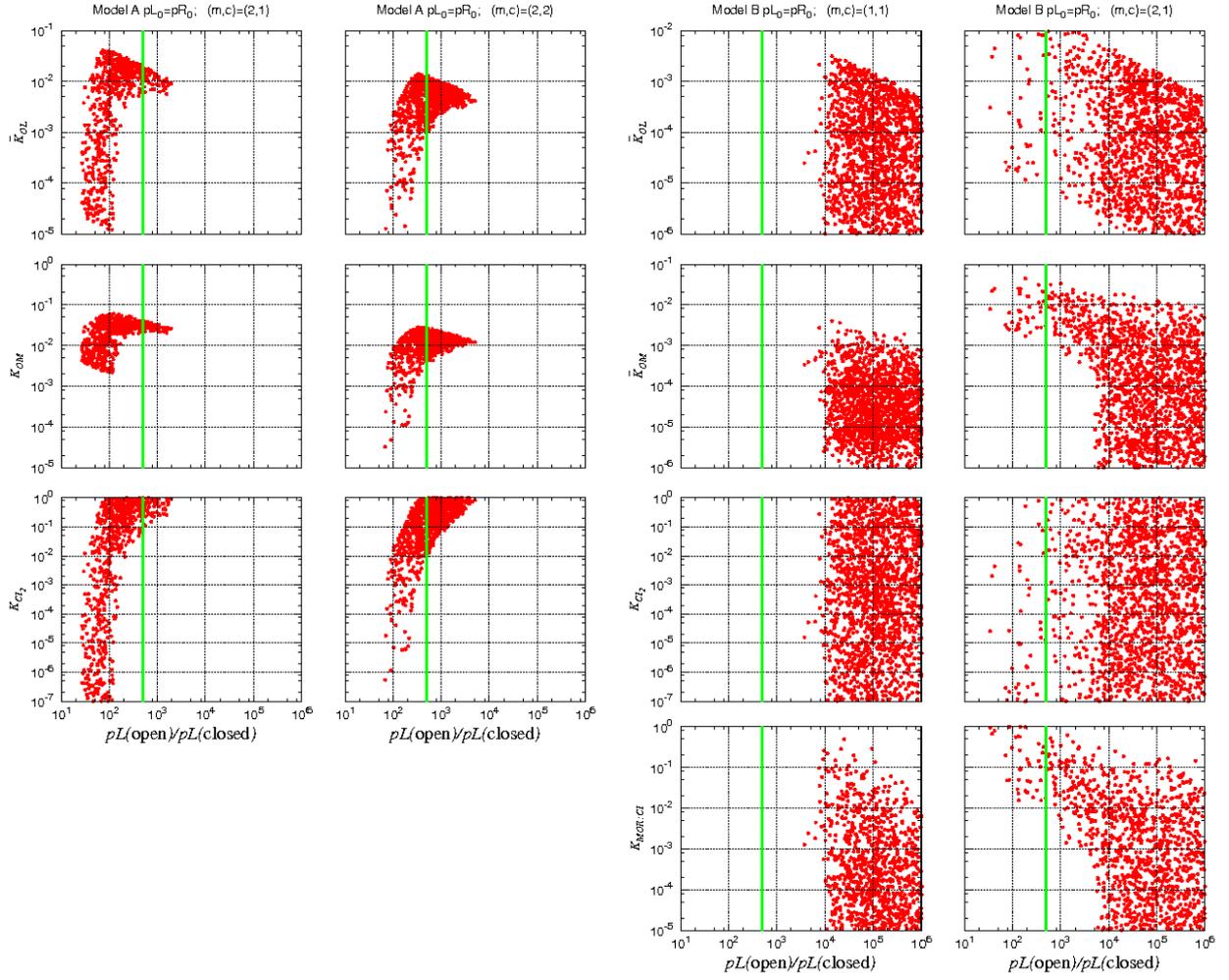}
\caption{Distribution of possible parameters
versus \PL repression fold, $pL({\rm open})/pL({\rm closed})$  
for Model A with $(m,c)$=(2,1) and (2,2), and
for Model B with $(m,c)$=(1,1) and (2,1).
The plotted parameters are those that satisfy the \PR repression fold
criterion $50< pR({\rm open})/pR({\rm closed})<200$ out of randomly
chosen parameters from the region $\tilde K_{\rm OL}, K_{\rm OM},
K_{\CI_2}\in [10^{-7}, 1]$ for Model A and $\tilde K_{\rm OL}, \tilde
K_{\rm OM}, K_{\CI_2}, K_{\MOR:\CI}\in [10^{-7}, 1]$ for Model B.
Vertical green lines are drawn at \PL repression fold 500, thus only the
parameters that are in the right side of the lines are consistent with
the experimentally obtained repression folds. 
The number of tested data are
$10^6$ for Model A with $(m,c)=(2,1)$,
$5\times 10^6$ for Model A with $(m,c)=(2,2)$, and
$2\times 10^5$ for Model B.
Note that the effective affinity $\tilde K_{\rm OM}$ for Model B is
 defined by 
$\tilde K_{\rm OM}^{m+c}\equiv K_{\rm OM} K_{\rm MOR_m CI_c}^{m+c-1}$.
}
\label{fig:ParamScan-Model-AB}
\end{figure}
}
\twocolumn

\begin{figure}[t]
\centerline{\includegraphics[width=0.5\textwidth]{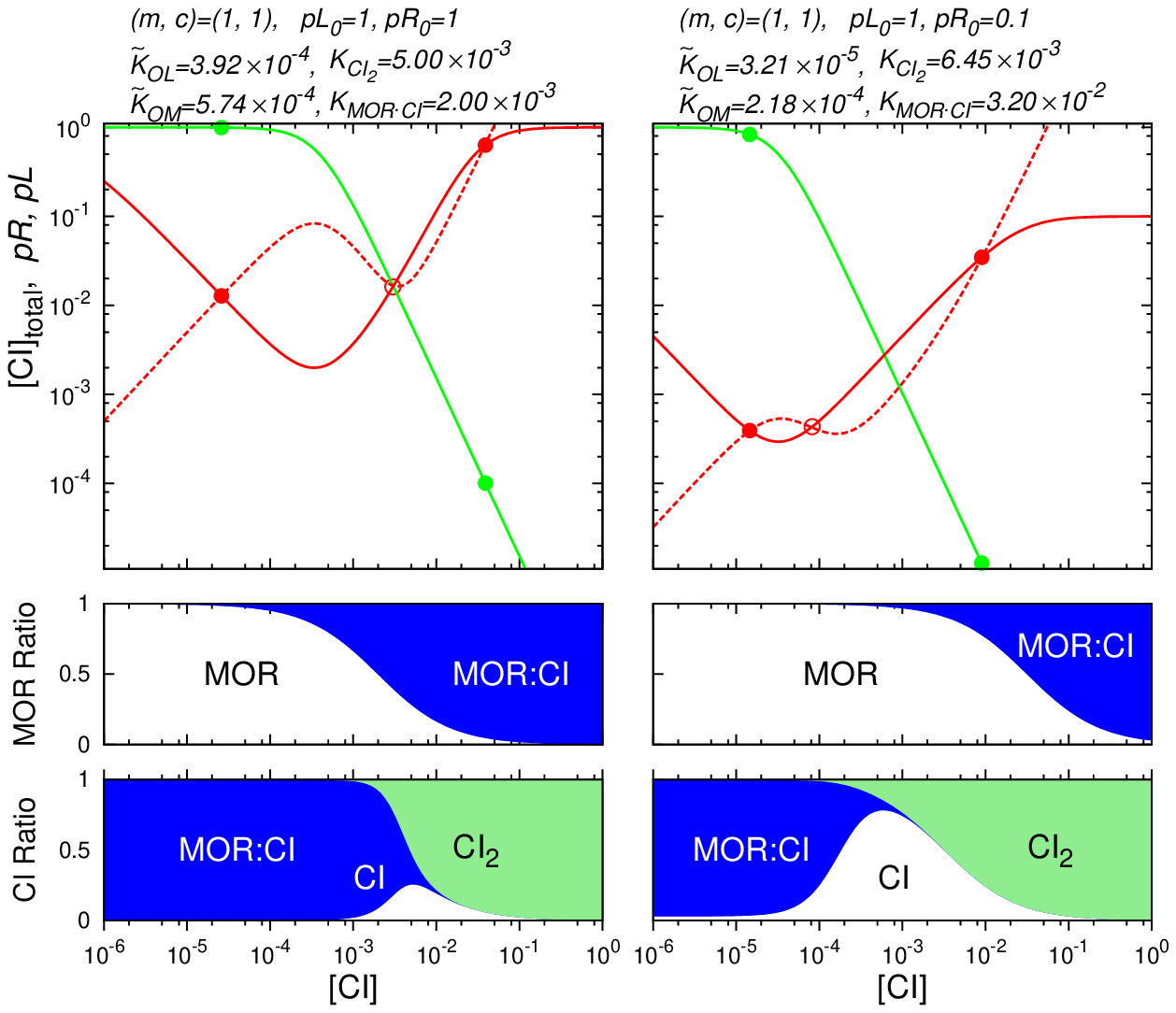}
}
\caption{ Model B with $\MOR\;\CI$ binding to \OMc, $(m,c)=(1,1)$. The bare
promoter activities, $pL_0$ and $pR_0$, are equal in left panel while
the bare activity $pL_0$ is 10 times stronger than $pR_0$ in the right
panel.
The upper graphs show examples of the promoter activities, $pL$ and
$pR$, and $[\CI]_{\rm total}$ as a function of [CI]. The repression
folds of \PL and \PR are approximately 1000 and 50, respectively in the
left figure, and approximately 60,000 and 100 in the right figure.
The middle graphs show the ratios of MOR units in the forms of monomer
and heteromer, and the lower graphs show the ratios of CI units in the
form of monomer, dimer, and heteromer.  One can see that MOR:CI and
$\CI_2$ compete for the free CI units in the intermediate concentration
range of [CI].  } \label{Model-B-m1c1}
\end{figure}
\begin{figure}[t]
\centerline{
\includegraphics[width=0.5\textwidth]{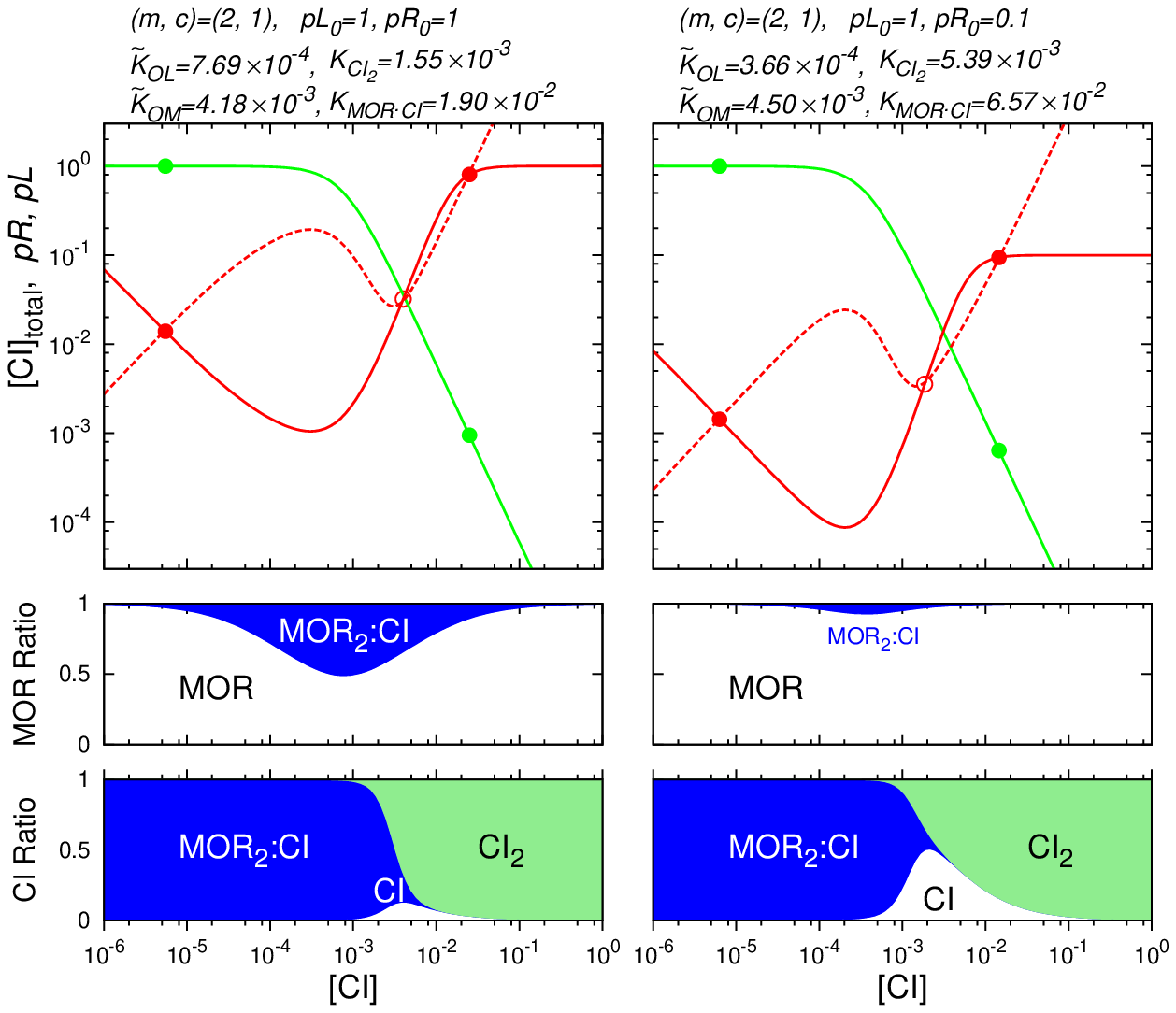}
}
\caption{Model B with $\MOR_2\CI$ binding to \OMc, $(m,c)=(2,1)$. The
bare promoter activities, $pL_0$ and $pR_0$, are equal in left panel
while the bare activity $pL_0$ is 10 times stronger than $pR_0$ in the
right panel.  The upper graphs show examples of the promoter activities,
$pL$ and $pR$, and $[\CI]_{\rm total}$ as a function of [CI]. The
repression folds of \PL and \PR are approximately 1,000 and 50,
respectively, in left figure, and approximately 1,000 and 60 in right
figure.
In contrast to the case with $(m,c)=(1,1)$ in Fig.\ref{Model-B-m1c1},
the heteromer $\MOR_2$:CI is not formed at high [CI] because
the \PL promoter is closed faster than the $\MOR_2$:CI is formed.
Note that the effective affinity of \OM is 
$\tilde K_{\rm OM}=(K_{\rm OM}K_{\rm MOR\cdot CI}^2)^{1/3}$.
} \label{Model-B-m2c1}
\end{figure}

We also examine Model B in the case where a larger complex,
$\MOR_m\CI_c$, represses \PRc.  Detailed formalism is given in the
appendix.

Figure \ref{Model-B-m2c1} shows the plots for the extended Model B with
$(m,c)=(2,1)$.  This version of the model shows sharper transition
between the $\MOR_2\CI$ regime and the $\CI_2$ regime for the form of CI
protein as one can see in the lower graphs for CI ratio.  As for the
form of MOR, substantial fraction of $\MOR_2\CI$ appears only in the
intermediate range of [CI] in contrast to the case of the $\MOR\,\CI$
heterodimer in Fig.\ref{Model-B-m1c1}.  This is because there are not
enough MOR's in the high [CI] region to form $\MOR_2\CI$ complex because
it requires two MOR proteins.

In the right two columns of Fig.\ref{fig:ParamScan-Model-AB}, the
parameters that give the \PR repression fold in the range [50, 200] are
plotted versus the resulting \PL repression fold for Model B.  The
$2\times 10^5$ parameter sets are chosen randomly over the range of
$[10^{-7},1]$ for $\tilde K_{\rm OL}$, $\tilde K_{\rm OM}$, $K_{\CI_2}$,
and $K_{\MOR:\CI}$.  One can see that broad range of parameter sets
satisfy the repression fold criterion for \PRc, but the resulting
repression folds for \PL are limited to the region larger than 5,000 for
$(m,c)=(1,1)$ and 50 for (2,1).

\section{Discussion}

\subsection{Summary}
The bacteriophage TP901-1 has provided us with a conceptually new design
of a genetic switch, in which the interaction between two
antagonistic regulators, CI and MOR, is essential.
Bistability between the immune and the ant-immune states has been
demonstrated with a genetic switch that consists of the two divergently
oriented promoters \PL and \PRc, the two promoter-proximal genes,
\textit{cI} and \textit{mor}, and only one of the three CI operator
sites \OL on a low copy number plasmid \cite{madsen1999, pedersen2008a}.
The repression folds in the two states have been determined by
\textit{in vivo} measurements as
around 1,000-fold for \PL repression in the immune state and
around 100-fold for \PR repression in the
anti-immune state.

We constructed mathematical models for this cloned bistable system,
assuming a putative operator \OM to regulate \PR
(Fig.\ref{fig:circuit}b).  We assumed that \PL is repressed by $\CI_2$
bound to \OL whereas \PR is repressed by the $\MOR_m\CI_c$:DNA complex
on \OM.  We examined two types of models:  one where MOR and CI
interact only on DNA (Model A), and the other where MOR and CI form
$\MOR_m\CI_c$ complex in cytoplasm first and then the complex binds to
\OM (Model B).  For each model, we tested bistability and performed
parameter scans using the criterion that the repression folds should be
consistent with the experiments.

\begin{table}
\begin{center}
\begin{tabular*}{0.5\textwidth}{@{\extracolsep{\fill}}lcc}
\hline \hline
 \multicolumn{3}{c}{Parameter Ranges for Model A}
\\
$(m,c)$ & (2,1) & (2,2) \\
\hline \\
$\tilde K_{\rm OL}$ &
$\sim 10^{-2}$ & $10^{-3} \sim 10^{-2}$ 
\\
$K_{\rm OM}$ &
$10^{-2}\sim 10^{-1}$  & $\sim 10^{-3}$
\\
$K_{\CI_2}$ &
$\gtrsim 10^{-2}$ & $\gtrsim 10^{-2}$
\\
\hline \hline
\end{tabular*}
\end{center}
\caption{Accepted ranges of parameters for Model A.  The values are
given in the unit of the concentrations that correspond to [CI] and
[MOR] at the full activity of the promoter \PR and \PLc, respectively.
} \label{T-1}
\end{table}
\begin{table}
\begin{center}
\begin{tabular*}{0.5\textwidth}{@{\extracolsep{\fill}}lcc}
\hline \hline
 \multicolumn{3}{c}{Parameter Ranges for Model B}
\\
$(m,c)$ & (1,1) & (2,1) \\
\hline \\
$\tilde K_{\rm OL}$ &
$\lesssim 3\times 10^{-3}$ & $(3\times 10^{-6} )\sim 3\times 10^{-2}$ 
\\
$\tilde K_{\rm OM}$ &
$\lesssim 5\times 10^{-3}$  & $(10^{-4})\sim 3\times 10^{-2}$
\\
$K_{\CI_2}$ &
--- & ---
\\
$K_{\MOR:\CI}$ &
$\lesssim 3\times 10^{-1}$ & $(10^{-3})\sim 1$
\\
\hline \hline
\end{tabular*}
\end{center}
\caption{Accepted ranges of parameters for Model B.  The values are
given in the unit of the concentrations that correspond to [CI] and
[MOR] at the full activity of the promoter \PR and \PLc, respectively.
The values in the parentheses are the lower limits when the fold
criterion for \PL is restricted to $500<pL({\rm open})/pL({\rm
closed})<5000$.  The ranges for $K_{\rm CI_2}$ cannot be set because the
accepted values extends over the whole tested range.  } \label{T-2}
\end{table}
Our results are summarized as follows:
For Model A, 
(i) the system shows bistability only when $2m-c\geq 2$,
(ii) the possible values for the operator affinities $\tilde K_{\rm OL}$
and $K_{\rm OM}$ are narrowly distributed,
%
%
(iii) the possible dissociation constant $K_{\CI_2}$ is
much larger than the operator affinity $\tilde K_{\rm OL}$ due to the large
repression fold for \PLc,
(iv) the possible value for $K_{\CI_2}$ is bounded by the relatively
large lower limit.
The accepted ranges for the parameters are listed in Table \ref{T-1}.
For Model B, 
(i) the bistability is robust due to the sequestration of CI by the
$\MOR_m\CI_c$ complex formation,
(ii) large parameter regions are allowed by the repression fold criterion,
(iii) the \PL repression fold is bounded by the lower limit for the
parameters that are consistent with the \PR repression fold:
$pL({\rm open})/pL({\rm closed}) > 5000$ for $(m,c)=(2,1)$, and
$> 50$ for $(m,c)=(2,2)$.
The accepted ranges for the parameters are listed in Table \ref{T-2}.
\\

\subsection{Validity of the Models}

In order to assess validity of the models,
we have to determine our unit for CI concentration first.
We employed the unit where [CI] is measured by the concentration in the
steady state with the full activity of \PLc.  This concentration should
be compared with [CI] in the immune state of {\it in vivo} experiment on
the system with around 10 copy-number plasmid containing the modified
switch.  We estimate this as follows by using the value of 300 nM for
the CI concentration in the lysogenic/immune state of the wild lambda
phage \cite{reichardt1971}; First, we assume that this concentration is
comparable with that for wild type TP901-1 with a single copy.  Then, we
multiply this by the following two factors: the factor 10 of the
copy-number of plasmid, and the factor 100 of the relative activity of
\PR in our modified system in comparison with the wild type switch
\cite{pedersen2008a}.  With these factors, we estimate that $pR_0$,
i.e., [CI] at the full \PR activity in the present system, could be well
over $10^{5}$ nM scale.

\paragraph{Model A}
With this unit,
we might be able to rule out Model A based upon the estimated values of
$\tilde K_{\rm OL}$ and $K_{\CI_2}$.
The possible value of $\tilde K_{\rm OL}$ in Model A is between around
$10^{-3}$ and $5\times 10^{-2}$ (Table \ref{T-1}), but this contradicts
the {\it in vitro} estimate of 28 nM for the CI concentration at which
\OL is occupied by CI for 50\% of the time\cite{johansen2003}.  The
lower limit of $\tilde K_{\rm OL}\sim 10^{-3}$ in Model A should
correspond to 100 nM or quite possibly even larger, but it is already
well above 28 nM, i.e., the {\it in vivo} estimate for repressor-DNA affinity
for TP901-1.

We also found that the large repression fold of \PL entails
$K_{\CI_2}\gg \tilde K_{\rm OL}$ for Model A.  This means that CI's
exist as monomers in cytoplasm and form $\CI_2$ when they bind to \OLc,
but this is in contrast with many phage-encoded repressor proteins, such
as those encoded by phage lambda, 434, and 186, which tend to exist as
dimers or higher oligomers in
solution\cite{Koblan1991,ciubotaru1999,neufing2001,Shearwin2002};
Actually, most of the 434 and lambda repressors exist in the dimeric
conformation at nanomolar concentrations \cite{neufing2001,Koblan1991}.
Our Model A challenges the presumption that the formation of dimers is a
prerequisite for its specific DNA binding.

\paragraph{Model B}
We found  broader distribution of parameter sets that satisfy the
repression fold criterion for Model B. In particular, we did not find
lower bounds for possible $K_{\CI_2}$ in contrast to the case of Model
A.

In the comparison of the two variants of Model B, our results show that
the model with the formation of $\MOR_2 \CI$ complex is more favorable
than that with $\MOR\,\CI$.  For the model with $\MOR\,\CI$, the \PL
repression fold turned out to be always larger than 5,000 for the
parameters that give the \PR repression fold between 50 through 200.
Such a high repression fold of \PL has never been actually observed {\it
in vivo}.  On the other hand, for the model with $\MOR_2\CI$, the lower
bound for the resulting \PL repression fold can be as low as 50, which
covers the observed range of the \PL repression fold.  As for the
parameter ranges, this variant of Model B gives $\tilde K_{\rm
OM}\gtrsim 0.2\times 10^{-4}$ and $K_{\rm MOR:CI}\gtrsim 10^{-3}$ when
the repression fold for \PL is smaller than 5,000
(Fig.\ref{fig:ParamScan-Model-AB}).


\subsection{Experimental Test}

\begin{figure}
\centerline{
\includegraphics[width=0.5\textwidth]{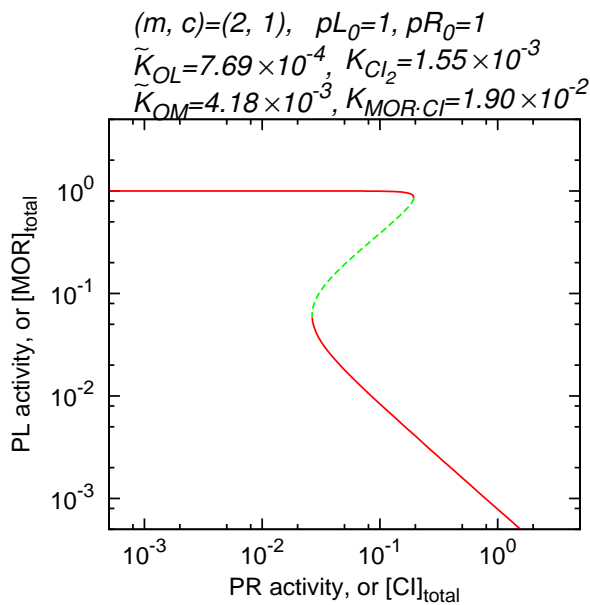}
}
\caption{
The \PL activity (or [MOR]$_{\rm total}$) versus
the \PR activity (or [CI]$_{\rm total}$) for the system with
 uncontrolled \PRc.
The solid lines show the stable states and the dashed line shows the
 unstable state.
}
\label{fig:UnCntl}
\end{figure}

One of the distinguishing consequences of Model B is that the system
with uncontrolled \PR can be bistable because of the sequestration of CI
by MOR:CI complex formation.  Even if \PR produces CI at a constant
rate, there can be the two stable states: one with the repressed \PL and
the other with the derepressed \PLc.  Such a mechanism of bistability
has been proposed by Fran\c{c}ois and Hakim
\cite{Francois2004,Francois2005} as a theoretical possibility.  Our
study suggests this mechanism is employed in TP901-1.  
This may be tested experimentally for the genetic switch of phage
TP901-1 by measuring the promoter activity of \PL in systems containing a
functional {\it mor} gene and expressing CI from uncontrolled \PR promoters at
constant but various rates.
Plotting the \PL activity of each system versus the uncontrolled \PR
activity, one should find a characteristic feature for bistability as in
Fig.\ref{fig:UnCntl}, where the \PL activity is doubled-valued for a
certain range of the \PR activity.  

\subsection{Concluding Remark}
The genetic switching mechanism in TP901-1 is remarkably robust; The
modified system studied here with only one operator \OL contains 100 times
more CI molecules in its immune state than the wild type genetic switch
with all of the three operators on plasmids, yet still shows bistability.  
Our model study suggests that the robustness of the genetic switch in
TP901-1 is brought about by sequestration of CI through MOR:CI
complex formation in cytoplasm.


\section*{Acknowledgements}
This work was supported by the Danish National Research
Foundation through the Center for Models of Life.


\section*{Appendix: Formalism for Model B with $(m,c)=(2,1)$}

In the appendix, we present some of the formulae we used for Model B
with $(m,c)=(2,1)$.
For this case, the promoter \PR activity is  a function of
$\MOR_2\CI$ concentration,
\begin{equation}
pR([\MOR], [\CI]) = pR_0{1\over 1+[\MOR_2\CI]/K_{\rm OM}}
\label{pR},
\end{equation}
and $[\MOR_2\CI]$ is given by 
\begin{equation}
[\MOR_2\CI] = {[\MOR]^2\cdot [\CI]\over K_{\MOR:\CI}^2},
\label{MOR_2CI}
\end{equation}
with the dissociation constant $K_{\MOR:\CI}$.

The total concentrations of CI and MOR are now
\begin{eqnarray}
[\CI]_{\rm total} &  = & [\CI] + 2[\CI_2] + [\MOR_2\CI],
\label{CI_total}
\\
{[\MOR]_{\rm total}} & = & [\MOR] + 2[\MOR_2\CI],
\label{MOR_total}
\end{eqnarray}
thus the corresponding equations with Eq.(\ref{MOR-B}) in the main
text becomes
\begin{eqnarray}
\lefteqn{
[\MOR] = {K_{\MOR :\CI}^2\over 4 [\CI]}\,
   \Biggl( -1} \nonumber \\
& & \quad
 + \sqrt{1+8 [\CI] [\MOR]_{\rm total}/K_{\MOR :\CI}^2 }\, \Biggr).
\label{MOR}
\end{eqnarray}
Thus, both sides of the steady state condition Eq.(\ref{SCE-pR})
are now given by
\begin{equation}
pR([\MOR],[\CI])  = 
 pR_0{1\over 1+[\MOR]^2[\CI]/\tilde K_{\rm OM}^3}
\end{equation}\begin{equation}
{[\CI]_{\rm total}}  =  
[\CI] + 2{[\CI]^2\over K_{\CI_2}} +
{1\over 2}\Bigl([\MOR]_{\rm total}-[\MOR]\Bigr)
\end{equation}
with
[MOR] by Eq.(\ref{MOR}) and $[\MOR]_{\rm total}$ by Eq.(\ref{SCE-pL}).
The effective affinity $\tilde K_{\rm OM}$ is defined as
\begin{equation}
\tilde K_{\rm OM}\equiv ( K_{\rm OM} K_{\MOR :\CI}^2)^{1/3}.
\end{equation}
%

\include{supplement}

\end{document}

%% file: supplement.tex
\def\mib#1{\mbox{\boldmath $#1$}}

\setcounter{section}{0}
\setcounter{equation}{0}
\setcounter{page}{1}
\onecolumn
\large

\begin{center}
{\Large\bf
Supplementary material for\\
``Model Analysis of the Genetic Switch Isolated from the
Temperate Bacteriophage TP901-1:\\
Repressor Sequestration Effect''
}
\vskip 2ex

{\bf Hiizu Nakanishi$^a$, Margit Pedersen$^b$, 
Anne K. Alsing$^b$, and Kim Sneppen$^b$}
\vskip 1ex

{\it
$^a$Department of Physics, Kyushu University 33, Fukuoka 812-8581, Japan \\
$^b$Niels Bohr Institute, Copenhagen University, Denmark
}
\end{center}
\vskip 3ex


\centerline{\parbox{13cm}{\small The stability criterion for the steady
state used in the manuscript is examined.  Dynamical analysis shows that
the criterion is valid as long as the $[\CI]_{\rm total}$ curve is a
monotonically increasing function.  In Model B, however, the $[\CI]_{\rm
total}$ curve has a part with negative slope for some parameter region,
in which case the stability cannot be determined only by comparing the
slopes of the curves.  The steady state at the left side could be
unstable when it is loacated in the region where $[\CI]_{\rm total}$ is
decreasing.  }}

\vskip 4ex


\section{Steady States}
\vskip 1ex

The steady states satisfy the self-consistent conditions
\begin{eqnarray}
pL\Bigl([CI]\Bigr) & = & [MOR]_{\rm tot}\Bigl([MOR],[CI]\Bigr) \label{SCE-MOR}
\\
pR\Bigl([MOR], [CI]\Bigr) & = & [CI]_{\rm tot}\Bigl([CI],[MOR]\Bigr),
\label{SCE-CI}
\end{eqnarray}
with
\begin{eqnarray}
[CI]_{\rm tot} \Bigl([CI],[MOR]\Bigr) & = & [CI] + 2[CI_2] + [MOR\cdot CI]
\label{CI_tot}
\\
{[MOR]_{\rm tot}}\Bigl([MOR],[CI]\Bigr) & = & [MOR] + [MOR\cdot CI] .
\label{MOR_tot}
\end{eqnarray}
Here, $[CI_2]$, $[MOR\cdot CI]$ are equilibrium concentrations of
the protein complexes.

We determined the steady solutions graphically by looking for
intersections of the following two curves as a function of $[CI]$,
i.e. the production curve and the $[\CI]_{\rm total}$ curve that
represents degradation rate:
\begin{eqnarray}
 pR & = &
 pR\Biggl([MOR],\, [CI]\Biggr)
\label{pR-HD}
\\
 {[CI]_{\rm tot}} & = &
 [CI]_{\rm tot}\Biggl([CI],[MOR] \Biggr),
\label{CI_t-HD}
\end{eqnarray}
with $[MOR]$  being a function
of $[MOR]_{\rm tot}$ and $[CI]$ derived from the relation $[MOR]_{\rm
tot}([MOR], [CI])$, and $[MOR]_{\rm tot}$ being given by $pL([CI])$,
\begin{equation}
[MOR] = [MOR]\Bigl([MOR]_{\rm tot},\, [CI]\Bigr)
=  [MOR]\Bigl(pL([CI]),\, [CI]\Bigr).
\end{equation}

\section{Stability criterion}
\vskip 1ex

In the text, the stability of the steady state is determined by the
simple criterion.
Let the slope of the production curve and the degradation curve be denoted by
\[
  {{\rm d}\, pR\over{\rm d} [CI]} \quad\mbox{and}\quad
  {{\rm d}\, [CI]_{\rm tot}\over{\rm d} [CI]},
\]
respectively.
Then the criterion is

\begin{equation}\begin{array}{rl}
\mbox{the steady state is stable }& \mbox{if }\displaystyle
  {{\rm d}\, pR\over{\rm d} [CI]} <  {{\rm d}\, [CI]_{\rm tot}\over{\rm d} [CI]},
\\ \\
\mbox{the steady state is unstable }& \mbox{if }\displaystyle
  {{\rm d}\, pR\over{\rm d} [CI]} >  {{\rm d}\, [CI]_{\rm tot}\over{\rm d}
  [CI]}
\label{criterion}
\end{array}\end{equation}
at the corresponding intersection.

This criterion is simple and plausible, but based on the single variable
picture although the system has at least two dynamical variables: $[CI]$
and $[MOR]$.  The full analysis for stability requires dynamical
consideration.

\section{Dynamical Analysis of Stability}
\vskip 2ex

The dynamics for the protein concentrations is given by the set of equations:
\begin{eqnarray}
{d\over dt}[MOR]_{\rm tot} & = &
{1\over \tau_M}\Bigl(pL([CI])-[MOR]_{\rm tot}\Bigr)
\label{MOR-t}
\\
{d\over dt}[CI]_{\rm tot} & = &
{1\over \tau_C}\Bigl(pR([MOR],[CI])-[CI]_{\rm tot}\Bigr).
\label{CI-t}
\end{eqnarray}
The total concentrations $[CI]_{\rm tot}$ and $[MOR]_{\rm tot}$ are
given by eqs.(\ref{CI_tot}) and (\ref{MOR_tot}).
Basic assumption for this is that the equilibration among protein
complexes in cytoplasm is much faster than the decay
rates of CI and MOR: $1/\tau_C$ and $1/\tau_M$.
\vskip 2ex

Consider the steady solution with $[CI]^*$ and $[MOR]^*$, which satisfies
eqs.(\ref{SCE-MOR}) and (\ref{SCE-CI}).
Suppose the steady state is perturbed by small fluctuation as
\begin{eqnarray}
[CI] & = & [CI]^* + \delta[CI]
\\
{[MOR]} & = & [MOR]^* + \delta[MOR],
\end{eqnarray}
and see if the small deviation will grow or decay in time.

By inserting these into eqs.(\ref{MOR-t}) and (\ref{CI-t}), we obtain
the equations for the time evolution of $\delta[CI]$ and $\delta[MOR]$,
\begin{eqnarray}
\lefteqn{
\left(\begin{array}{cc}  \displaystyle
\left(\partial M_t\over \partial C\right)^*, &\displaystyle
\left(\partial M_t\over \partial M\right)^*
\\ \\\displaystyle
\left(\partial C_t\over \partial C\right)^*, &\displaystyle
\left(\partial C_t\over \partial M\right)^*
\end{array}\right)
\left(\begin{array}{c}
\delta\dot C \\ \\ \delta\dot M
\end{array}\right)
=
}\nonumber \\ \nonumber \\ & & \qquad
\left(\begin{array}{cc}\displaystyle
{1\over\tau_M}\left\{\left({\partial L\over\partial C}\right)^*-
               \left({\partial M_t\over\partial C}\right)^*\right\},
&\displaystyle
-{1\over\tau_M}\left({\partial M_t\over\partial M}\right)^*
\\ \\ \displaystyle
{1\over\tau_C}\left\{\left({\partial R\over\partial C}\right)^*-
               \left({\partial C_t\over\partial C}\right)^*\right\},
&\displaystyle
{1\over\tau_C}\left\{\left({\partial R\over\partial M}\right)^*-
               \left({\partial C_t\over\partial M}\right)^*\right\}
\end{array}\right)
\left(\begin{array}{c}
\delta C \\ \\ \delta M
\end{array}\right),
\label{PT-eq-0}
\end{eqnarray}
where we employ the abbreviated notations:
\[
 M_t \equiv [MOR]_{\rm tot}, \quad
 M \equiv [MOR], \quad
 C_t \equiv [CI]_{\rm tot}, \quad
 C \equiv [CI],  \quad
 L\equiv pL, \quad   
 R\equiv pR.
\]
We further abbreviate the notation as
\[
 M_{t,C}\equiv \left({\partial M_t\over\partial C}\right)^*
= \left({\partial [MOR]_{\rm tot}\over\partial [CI]}\right)^*, \quad
 L_{,C} \equiv \left({\partial L\over\partial C}\right)^*
=\left({\partial\, pL([CI])\over\partial [CI]}\right)^*, \quad
\mbox{etc.}
\]
then eq.(\ref{PT-eq-0}) is expressed as
\begin{equation}
\left(\begin{array}{cc}  \displaystyle
M_{t,C}\, , & M_{t,M}
\\ \\
C_{t, C}\, , & C_{t, M}
\end{array}\right)
\left(\begin{array}{c}
   \delta\dot C \\ \\ \delta\dot M
\end{array}\right)
=
\left(\begin{array}{cc}\displaystyle
{1\over\tau_M}\Bigl(L_{,C}- M_{t, C}\Bigr),
&\displaystyle
-{1\over\tau_M} M_{t, M}
\\ \\ \displaystyle
{1\over\tau_C}\Bigl( R_{,C} - C_{t, C} \Bigr),
&\displaystyle
{1\over\tau_C}\Bigl( R_{, M} -  C_{t, M} \Bigr)
\end{array}\right)
\left(\begin{array}{c}
\delta C \\ \\ \delta M
\end{array}\right).
\label{PT-eq}
\end{equation}


\section
{The criterion is always valid for Model A:}
\vskip 2ex

In the case of Model A, $[MOR\cdot CI]=0$, thus we have
\[
 C_{t,M} = 0, \quad  M_{t, M} = 1, \quad  M_{t, C} = 0,
\]
then eq.(\ref{PT-eq}) becomes
\begin{equation}
\left(\begin{array}{cc}
0, & 1
\\ \\
C_{t, C}, & 0
\end{array}\right)
\left(\begin{array}{c}
\delta\dot C \\ \delta\dot M
\end{array}\right)
=
\left(\begin{array}{cc}
{1\over\tau_M}L_{, C}\, ,
&
-{1\over\tau_M}
\\ \\
{1\over\tau_C}\Bigl( R_{, C} - C_{t, C} \Bigr),
&
{1\over\tau_C}R_{, M}
\end{array}\right)
\left(\begin{array}{c}
\delta C \\ \delta M
\end{array}\right).
\end{equation}

Now we assume the solution as
\[
 \delta C, \quad \delta M \propto e^{\omega t}
\]
then we have
\begin{equation}
\left(\begin{array}{cc}
{1\over\tau_M}L_{, C},
&
-{1\over\tau_M} - \omega
\\ \\
{1\over\tau_C}\Bigl( R_{, C} - C_{t, C}\Bigr)
-C_{t, C}\omega,
&
{1\over\tau_C}R_{, M}
\end{array}\right)
\left(\begin{array}{c}
\delta C \\ \delta M
\end{array}\right)
=0.
\end{equation}

The condition that this equation has non-zero solution gives
\begin{equation}
 -C_{t, C}\, \omega^2 +
\left[
   {1\over\tau_C}\Bigl(R_{, C} - C_{t, C}\Bigr) - {1\over\tau_M}C_{t, C}
\right]\omega
+
{1\over\tau_M\tau_C}\Bigl[  L_{, C} R_{, M} + R_{, C} - C_{t, C} \Bigr]
=0
\label{omega-eq-Mono}
\end{equation}

If all the solutions $\omega$ have a negative real part,
the state is stable, whereas the state is unstable if there is a
solution with a positive real part.
\vskip 2ex

Note that the slopes of the production curve and the degradation curve are
given by
\begin{eqnarray}
{{\rm d}\, pR\over{\rm d}CI} & \equiv &
 \left({ {\rm d}\, pR([MOR],[CI])\over {\rm d}[CI] }\right)^* 
 =    L_{, C} R_{, M} + R_{, C}
\\
{{\rm d} CI_{\rm tot}\over{\rm d} CI} & \equiv &
\left({\partial [CI]_{\rm tot}([CI])\over\partial [CI]} \right)^*
 = C_{t,C}.
\end{eqnarray}

Using these, eq.(\ref{omega-eq-Mono}) becomes
\begin{eqnarray}\lefteqn{
 -{{\rm d} CI_{\rm tot}\over{\rm d}CI}\, \omega^2 
+ \left[
{1\over\tau_C}\left(
   {{\rm d}\, pR\over{\rm d}CI}-{{\rm d} CI_{\rm tot}\over{\rm d}CI}
       - L_{, C}R_{, M}
             \right)
  -
{1\over\tau_M}{{\rm d} CI_{\rm tot}\over{\rm d}CI}
\right] \omega
}\hspace{7cm}\nonumber\\ & &
+ {1\over\tau_C\tau_M}\left[
   {{\rm d}\, pR\over{\rm d}CI}-{{\rm d} CI_{\rm tot}\over{\rm d}CI}
  \right]
=0
\label{omega-eq-Mono-2}
\end{eqnarray}

Note that
\[
   L_{, C},\quad R_{, C},\quad R_{, M} < 0.
\]

\paragraph{(1) The case of $\tau_M\ll\tau_C$}
One solution is of order  $1/\tau_M$ and the other is of order $1/\tau_C$.
\[
   \omega \approx \left\{\begin{array}{l}\displaystyle
-{1\over\tau_M}
\\\\\displaystyle
{1\over\tau_C}\left[{{\rm d} CI_{\rm tot}\over{\rm d}CI}\right]^{-1}
\left[
   {{\rm d}\, pR\over{\rm d}CI}-{{\rm d} CI_{\rm tot}\over{\rm d}CI}
\right]
\end{array}\right. .
\]
This shows the criterion (\ref{criterion}) is valid.

\paragraph{(2) The case of $\tau_C\ll\tau_M$}
One solution is of order  $1/\tau_M$ and the other is of order $1/\tau_C$.
\[
   \omega \approx \left\{\begin{array}{l}\displaystyle
{1\over\tau_C}\left[{{\rm d} CI_{\rm tot}\over{\rm d}CI}\right]^{-1}
\left[
   {{\rm d}\, pR\over{\rm d}CI}-{{\rm d} CI_{\rm tot}\over{\rm d}CI}
       - L_{, C}R_{, M}
\right]
\\\\\displaystyle
-{1\over\tau_M}
\left[
   {{\rm d}\, pR\over{\rm d}CI}-{{\rm d} CI_{\rm tot}\over{\rm d}CI}
       - L_{, C}R_{, M}
\right]^{-1}
\left[
   {{\rm d}\, pR\over{\rm d}CI}-{{\rm d} CI_{\rm tot}\over{\rm d}CI}
\right]
\end{array}\right. .
\]
The criterion (\ref{criterion}) is also valid because
\[
{{\rm d} CI_{\rm tot}\over{\rm d}CI}>0, \quad
L_{, C} R_{, M} >0 .
\]
\vskip 4ex

\paragraph{(3) General case}
The criterion (\ref{criterion}) can be shown to be valid
because eq.(\ref{omega-eq-Mono}) always has two real solutions,
$\omega_1$ and $\omega_2$,
and the sum of the two solutions is negative when
the slope of the production curve is less steep than that of the
degradation curve, i.e.
\[
   \omega_1+\omega_2<0 \quad\mbox{when}\quad
   {{\rm d}\, pR\over{\rm d}CI}-{{\rm d} CI_{\rm tot}\over{\rm d}CI}<0 . 
\]

\vskip 4ex
\begin{quote}
{\it The positivity of the discriminant $D$ of  eq.(\ref{omega-eq-Mono}):}
\begin{eqnarray*}
\lefteqn{D=
\left[
   {1\over\tau_C}\left\{R_{, C}- C_{t, C}\right\} -{1\over\tau_M}C_{t, C}
\right]^2
}
\\
& & \qquad
+
4C_{t, C}{1\over\tau_M\tau_C}\left[ L_{, C} R_{, M} + R_{, C} - C_{t, C}\right]
\\
& = &
\left[
{1\over\tau_C}R_{, C} -\left({1\over\tau_C}-{1\over\tau_M}\right) C_{t, C} 
\right]^2
+{4\over\tau_C\tau_M} C_{t, C} R_{, M} L_{, C} 
 \, > 0,
\end{eqnarray*}
thus eq.(\ref{omega-eq-Mono}) has two real solutions.
\end{quote}
%

\section{The criterion is valid for  Model B
as long as the degradation curve has pisitive slope, but the state may be
unstable otherwise.}
\vskip 2ex

In this model, the existence of $MOR\cdot CI$ makes the expressions for
the slopes of the production curve and the degradation curve a bit more
complicated:
\begin{eqnarray}
\lefteqn{
{{\rm d}\, pR\over{\rm d}CI}\equiv
{{\rm d}\over{\rm d}[CI] }\,
 pR\Biggl([MOR]\Bigl(M_t, [CI]\Bigr),\, [CI]\Biggr)
}\hspace{3cm}
\nonumber \\
& = &
\left(\partial R\over\partial C\right)_M
+
\left(\partial R\over\partial M\right)_C \left[
\left({\partial M\over\partial M_t}\right)_C
\left({\partial L\over\partial C}\right)
+
\left({\partial M\over\partial C}\right)_{M_t}
\right]
\nonumber \\
& = &
\left(\partial R\over\partial C\right)_M
+
\left(\partial R\over\partial M\right)_C 
\left({\partial M_t\over\partial M}\right)_C^{-1}
\left[
\left({\partial L\over\partial C}\right)
-
\left({\partial M_t\over\partial C}\right)_M
\right]
\nonumber \\
& = &
R_{,C} + {R_{,M}\over M_{t,M}}\Bigl(L_{,C}-M_{t,C}\Bigr)
\label{pR-slope-HD}
\\
\lefteqn{
{{\rm d}\, CI_{\rm tot}\over{\rm d}CI}\equiv
{{\rm d}\over{\rm d}[CI] }\,
 [CI]_{\rm tot}\Biggl([CI],[MOR]\Bigl(M_t,\,[CI]\Bigr) \Biggr)
}\hspace{3cm}
\nonumber \\
& = &
\left({\partial C_t\over\partial C}\right)_M
+
\left({\partial C_t\over\partial M}\right)_C 
\left[
  \left({\partial M\over\partial M_t}\right)_C
  \left({\partial L\over\partial C}\right)
  +
  \left({\partial M\over\partial C}\right)_{M_t}
\right]
\nonumber\\
& = & 
\left({\partial C_t\over\partial C}\right)_M
+
\left({\partial C_t\over\partial M}\right)_C 
  \left({\partial M_t\over\partial M}\right)_C^{-1}
\left[
  \left({\partial L\over\partial C}\right)
  -
  \left({\partial M_t\over\partial C}\right)_M
\right]
\nonumber\\
& = &
C_{t,C} + {C_{t,M}\over M_{t,M}}\Bigl( L_{,C} - M_{t,C}\Bigr),
\label{C_t-slope-HD}
\end{eqnarray}
where the variables that kept constant upon partial differentiation are
explicitly indicated as
\[
 \left({\partial M\over\partial M_t}\right)_C
 \equiv 
 {\partial\over\partial [MOR]_{\rm tot}}[MOR]\Bigl([MOR]_{\rm tot}, [CI]\Bigr)
\]
whenever it could be ambiguous.  In the derivation, we have used the
relations
\[
 \left({\partial M\over\partial M_t}\right)_C = 
    \left({\partial M_t\over\partial M}\right)_C^{-1}, \qquad
 \left({\partial M\over\partial C}\right)_{M_t}
 \left({\partial M_t\over\partial M}\right)_C
 \left({\partial C\over\partial M_t}\right)_M = -1.
\]

\subsection{Stability of the system with PL with externally controlled 
$[CI]_{\rm tot}$ 
}

First, we will examine the stability of the system without PR, CI being
provided externally.  The system is shown to be bistable for some
parameter region.

Based upon the approximation that the relaxation in
the solution is much faster than the protein production rate by PL, we
consider the system where $[CI]$ and $[MOR]$ satisfy
\begin{eqnarray}
{[CI]_{\rm tot}} & = &
 [CI]_{\rm tot}\Bigl([CI],[MOR] \Bigr)
\\
{d\over dt}[MOR]_{\rm tot} & = &
{1\over \tau_M}\Bigl(pL([CI])-[MOR]_{\rm tot}\Bigr),
\end{eqnarray}
thus the deviation from the steady state follows
\begin{eqnarray}
C_{t,C}\delta C + C_{t,M}\delta M & = & 0
\\
M_{t,C}\delta\dot C + M_{t,M}\delta\dot M & = &
{1\over\tau_M}\Bigl(
L_{,C}\delta C - M_{t,C}\delta C - M_{t,M}\delta M
\Bigr),
\end{eqnarray}
which results in
\begin{eqnarray}
\delta\dot C  & = &
{1\over\tau_M}\,{
      (L_{,C}-M_{t,C})C_{t,M}+M_{t,M}C_{t,C}  \over
      M_{t,C}C_{t,M}-M_{t,M}C_{t,C}
                }\,\,\delta C
\nonumber \\
& = &
-{1\over\tau_M}\,{ M_{t,M}\over A}\,\,
       {{\rm d}CI_{\rm tot}\over{\rm d}CI} \,\,\delta C,
\end{eqnarray}
with the notation
\[
      A\equiv C_{t,C} M_{t,M} - C_{t,M} M_{t,C} >0,
\]
whose inequality can be shown
from the actual expressions of $[CI]_{\rm tot}$ and $[MOR]_{\rm tot}$,
(\ref{CI_tot}) and (\ref{MOR_tot}).

Therefore, we have
\begin{eqnarray*}
    {{\rm d}CI_{\rm tot}\over{\rm d}CI} <0 & & \mbox{unstable}\\ \\
    {{\rm d}CI_{\rm tot}\over{\rm d}CI} >0 & & \mbox{stable} .
\end{eqnarray*}


\subsection{Stability of genetic switch with PL and PR:}

The growth rate $\omega$ is determined by the characteristic equation
\begin{equation}
\left|\begin{array}{cc}
{1\over\tau_M}(L_{,C}-M_{t,C})-M_{t,C}\,\omega, &
-{1\over\tau_M}M_{t,M} - M_{t,M}\,\omega
\\ \\
{1\over\tau_C}(R_{,C}-C_{t,C})-C_{t,C}\,\omega, &
{1\over\tau_C}(R_{,M}-C_{t,M})-C_{t,M}\,\omega
\end{array}\right|
=0 ,
\label{det-HD}
\end{equation}
which can be expanded as
\begin{eqnarray}
&
\omega^2 \Bigl[ M_{t,C}C_{t,M}-C_{t,C}M_{t,M} \Bigr] 
\hspace{8cm}&
\nonumber \\
%
&\displaystyle
+\omega\Biggl[
{1\over\tau_C}\Bigl(-M_{t,C}(R_{,M}-C_{t,M})+M_{t,M}(R_{,C}-C_{t,C})\Bigr)
-
{1\over\tau_M}\Bigl( C_{t,M}(L_{,C}-M_{t,C})+C_{t,C}M_{t,M} \Bigr)
\Biggr]
&
\nonumber \\
&\displaystyle\hspace{1cm}
+{1\over\tau_M\tau_C}\Bigl[
(L_{,C}-M_{t,C})(R_{,M}-C_{t,M}) + (R_{,C}-C_{t,C})M_{t,M}
\Bigr]
 = 0. &
\label{disc-HD}
\end{eqnarray}
This can be put in the form
\begin{eqnarray}&\displaystyle
%
-\omega^2 {A\over M_{t,M}}
+
\omega \left[ 
\Biggl\{
{1\over\tau_C}
 \left({{\rm d}\,pR\over{\rm d}CI}-{{\rm d}\,CI_{\rm tot}\over{\rm d}CI}\right)
-
{1\over\tau_M}{{\rm d}\,CI_{\rm tot}\over{\rm d}CI}
\Biggr\}
-
{1\over\tau_C}{L_{,C}\Bigl( R_{,M} - C_{t,M} \Bigr)\over M_{t,M}}
\right] 
&
\nonumber \\ &\displaystyle
+
{1\over\tau_M\tau_C} \left[
    {{\rm d}\,pR\over{\rm d}CI}-{{\rm d}\,CI_{\rm tot}\over{\rm d}CI}
    \right]
=0 . &
\label{omega-eq-Hetero-2}
\end{eqnarray}
This is almost the same with the corresponding equation for
Model A eq.(\ref{omega-eq-Mono-2}).
\vskip 4ex

\paragraph{(1) The case of $\tau_M\ll\tau_C$}
One solution is of order $1/\tau_M$ and the other of order $1/\tau_C$:
\begin{equation}
\omega \approx \left\{ \begin{array}{l}\displaystyle
-{1\over\tau_M}\,{M_{t,M}\over A}
\left( {{\rm d}CI_{\rm tot}\over{\rm d}CI}\right)
\\\\\displaystyle
{1\over\tau_C}
\left[
 {{\rm d}\, pR\over{\rm d}CI} - {{\rm d}\, CI_{\rm tot}\over{\rm d}CI}
\right]
\left({{\rm d} CI_{\rm tot}\over{\rm d}CI}\right)^{-1}
\end{array}\right.  .
\end{equation}
Therefore, the criterion (\ref{criterion}) is valid as long as the slope
of degradation curve is positive, i.e. $({\rm d}CI_{\rm tot}/{\rm d}CI)>0$,
but the state is always unstable for the negative slope for the degradation
curve, i.e. $({\rm d}CI_{\rm tot}/{\rm d}CI)<0$.

\paragraph{(2) The case of $\tau_C\ll\tau_M$}
\begin{equation}
\omega \approx\left\{ \begin{array}{l}\displaystyle
{1\over\tau_C}\,{ M_{t,M}\over A}
\left[
 \left({{\rm d}\,pR\over{\rm d}CI}-{{\rm d}\,CI_{\rm tot}\over{\rm d}CI}\right)
         - {L_{,C}\Bigl( R_{,M}- C_{t,M} \Bigr)\over  M_{t,M}}
\right]
\\\\\displaystyle
-{1\over\tau_M}
\left[ 
 \left({{\rm d}\,pR\over{\rm d}CI}-{{\rm d}\,CI_{\rm tot}\over{\rm d}CI}\right)
         - {L_{,C}\Bigl( R_{,M} - C_{t,M} \Bigr) \over M_{t,M}}
\right]^{-1}
\left[
    {{\rm d}\,pR\over{\rm d}CI}-{{\rm d}\,CI_{\rm tot}\over{\rm d}CI}
    \right]
\end{array}\right.
\end{equation}

In this case, the criterion (\ref{criterion}) is always valid because
\[
  {L_{,C}\Bigl( R_{,M} - C_{t,M} \Bigr) \over M_{t,M}} >0, \qquad
  { M_{t,M}\over A} >0.
\]

\paragraph{(3) General case} 
We can show eq.(\ref{omega-eq-Hetero-2}) always has two real solutions,
$\omega_1$ and $\omega_2$, and the steady state stability is determined
from the sign of $\omega_1+\omega_2$:
\begin{eqnarray*}
{{\rm d}\,pR\over{\rm d}CI}-{{\rm d}\,CI_{\rm tot}\over{\rm d}CI}<0,
\quad &\displaystyle
{{\rm d}\,CI_{\rm tot}\over{\rm d}CI}>0
& \mbox{then  stable}
\\\\
{{\rm d}\,pR\over{\rm d}CI}-{{\rm d}\,CI_{\rm tot}\over{\rm d}CI}>0, \quad &
& \mbox{then  unstable}
\end{eqnarray*}
but
\begin{eqnarray*}
{{\rm d}\,pR\over{\rm d}CI}-{{\rm d}\,CI_{\rm tot}\over{\rm d}CI}<0,
\quad &\displaystyle
{{\rm d}\,CI_{\rm tot}\over{\rm d}CI}<0
& \mbox{  stable/unstable} .
\end{eqnarray*}

This means that {\em the stability cannot be determined only from the slopes
of the production and the degradation curves in the case the degradation curve
has a negative slope.}

\section*{Appendix: 
The positivity of the discriminant $D$ of eq. (\ref{disc-HD}):}

\begin{eqnarray*}
\displaystyle
D & =  &
\Biggl[
{1\over\tau_C}\Biggl(-M_{t,C}(R_{,M}-C_{t,M})+M_{t,M}(R_{,C}-C_{t,C})\Biggr)
\\ & & \hspace{4cm}
-
{1\over\tau_M}\Biggl( C_{t,M}(L_{,C}-M_{t,C})+C_{t,C}M_{t,M} \Biggr)
\Biggr]^2
 \\
& &
- {4\over\tau_M\tau_C}
\Biggl[ M_{t,C}C_{t,M}-C_{t,C}M_{t,M} \Biggr] 
\Biggl[
(L_{,C}-M_{t,C})(R_{,M}-C_{t,M}) + (R_{,C}-C_{t,C})M_{t,M}
\Biggr]
\\
& = &
\Biggl[{1\over\tau_C}\Biggl(
-A -R_{,M}M_{t,C} + R_{,C}M_{t,M} 
\Biggr)
-
{1\over\tau_M}\Biggl( A +L_{,C}C_{t,M}\Biggr)
\Biggr]^2
\\ & & \qquad\qquad
+{4\over\tau_M\tau_C}\, A\Biggl[
-A + L_{,C}R_{,M} - L_{,C} C_{t,M} -R_{,M}M_{t,C} + R_{,C}M_{t,M}
\Biggr]
\\ 
& = &
\Biggl[
   -A\Biggl({1\over\tau_C}+{1\over\tau_M}\Biggr)
   +\Biggl(  {1\over\tau_C}( -R_{,M}M_{t,C} + R_{,C}M_{t,M} )
           - {1\over\tau_M}L_{,C}C_{t,M}   \Biggr)
\Biggr]^2
\\ & & \qquad\qquad
+{4\over\tau_M\tau_C}\, A\Biggl[
-A + L_{,C}R_{,M} - L_{,C} C_{t,M} -R_{,M}M_{t,C} + R_{,C}M_{t,M}
\Biggr]
\\
& = &
A^2\left[ {1\over\tau_C}-{1\over\tau_M} \right]^2
+
2A \Biggl[
\Biggl({1\over\tau_C}+{1\over\tau_M}\Biggr)
 \Biggl(  {1\over\tau_C}( R_{,M}M_{t,C} - R_{,C}M_{t,M} )
           + {1\over\tau_M}L_{,C}C_{t,M}   \Biggr)
\\ & & \hspace{5cm}
+{2\over\tau_M\tau_C}\Biggl(
L_{,C}R_{,M} - L_{,C} C_{t,M} -R_{,M}M_{t,C} + R_{,C}M_{t,M}
\Biggr)
\Biggr]
\\  & & \qquad
   +\Biggl(  {1\over\tau_C}( -R_{,M}M_{t,C} + R_{,C}M_{t,M} )
           - {1\over\tau_M}L_{,C}C_{t,M}   \Biggr)^2
\end{eqnarray*}
The coefficient of $2A$ is
\begin{eqnarray*}
\lefteqn{
\Bigl( R_{,M}M_{t,C} - R_{,C}M_{t,M} \Bigr)
       \Biggl({1\over\tau_C}-{1\over\tau_M}\Biggr){1\over\tau_C}
+L_{,C}C_{t,M} \Biggl(-{1\over\tau_C}+{1\over\tau_M}\Biggr){1\over\tau_M}
+
L_{,C}R_{,M}{2\over\tau_M\tau_C},
}
\\
& = &
\Biggl({1\over\tau_C}-{1\over\tau_M}\Biggr)\left(
{1\over\tau_C}( R_{,M}M_{t,C} - R_{,C}M_{t,M} )
-
{1\over\tau_M}L_{,C}C_{t,M}
\right)
+
{2\over\tau_M\tau_C}L_{,C}R_{,M}.
\end{eqnarray*}
Thus, $D$ is
\begin{eqnarray*}
D 
& = &
A^2\left( {1\over\tau_C}-{1\over\tau_M} \right)^2
\\ & &
+
2A \Biggl[
\left({1\over\tau_C}-{1\over\tau_M}\right)\left(
{1\over\tau_C}( R_{,M}M_{t,C} - R_{,C}M_{t,M} )-
             {1\over\tau_M}L_{,C}C_{t,M}
\right)
+
{2\over\tau_M\tau_C}L_{,C}R_{,M}
\Biggr]
\\  & & \qquad
   +\Biggl(  {1\over\tau_C}( R_{,M}M_{t,C} - R_{,C}M_{t,M} )
           + {1\over\tau_M}L_{,C}C_{t,M}   \Biggr)^2
\\ 
& = &
\left[
\left({1\over\tau_C}-{1\over\tau_M}\right)A
+
\left( {1\over\tau_C}( R_{,M}M_{t,C} - R_{,C}M_{t,M} )
                   -{1\over\tau_M}L_{,C}C_{t,M} \right)
\right]^2
\\ & & \quad
-\left( {1\over\tau_C}( R_{,M}M_{t,C} - R_{,C}M_{t,M} )
                   -{1\over\tau_M}L_{,C}C_{t,M} \right)^2
 + 2A {2\over\tau_M\tau_C}L_{,C}R_{,M}
\\ & & \qquad
   +\Biggl(  {1\over\tau_C}( R_{,M}M_{t,C} - R_{,C}M_{t,M} )
           + {1\over\tau_M}L_{,C}C_{t,M}   \Biggr)^2
\\ 
& = &
\left[
\left({1\over\tau_C}-{1\over\tau_M}\right)A
+
\left( {1\over\tau_C}( R_{,M}M_{t,C} - R_{,C}M_{t,M} )
                   -{1\over\tau_M}L_{,C}C_{t,M} \right)
\right]^2
\\ & & \quad
+{4\over\tau_C\tau_M}L_{,C} \Biggl(
\Bigl( R_{,M}M_{t,C} - R_{,C}M_{t,M} \Bigr) C_{t,M} + 
\Bigl(C_{t,C}M_{t,M}-C_{t,M}M_{t,C}\Bigr) R_{,M}
\Biggr)
\end{eqnarray*}

\vskip 4ex
The second line of the last expression is shown to be positive as
follows:
\vskip 2ex

Note that
\[
C_{t,C}  = 1 + C_{2,C} + MC_{,C}, \quad
C_{t,M}  = MC_{,M}, \quad
M_{t,C}  = MC_{,C}, \quad
M_{t,M}  = 1 + MC_{,C},
\]\[
R_{,C} = R'\cdot MC_{,C}, \quad
R_{,M} = R'\cdot  MC_{,M},
\]
where the abbreviated notations are used,
\[
 C_{2,C} \equiv {\partial [CI_2]\over \partial [CI]}, \quad
 MC_{,M} \equiv {\partial [MOR\cdot CI]\over\partial [MOR]},
\quad \mbox{etc. and}\quad
 R' \equiv {\partial\, pR\over \partial [MOR\cdot CI]}.
\]

Then the second line of $D$ is
\begin{eqnarray*}
\lefteqn{ L_{,C}\Biggl(
\Bigl( R_{,M}M_{t,C} - R_{,C}M_{t,M} \Bigr) C_{t,M} + 
\Bigl(C_{t,C}M_{t,M}-C_{t,M}M_{t,C}\Bigr) R_{,M}        \Biggr)
}
\\
& = &
 L_{,C}\Bigl(  - R_{,C}M_{t,M} C_{t,M} + C_{t,C}M_{t,M} R_{,M} \Bigr)
\\
& = &
L_{,C}\, M_{t,M}( - R_{,C} C_{t,M} + R_{,M} C_{t,C} )
\\
& = &
L_{,C}\, M_{t,M}\Bigl( 
- R'\cdot MC_{,C} MC_{,M} + R'\cdot MC_{,M} (1+C_{2,C}+MC_{,C} )
\Bigr)
\\
& = &
L_{,C}\, R'\cdot MC_{,M} (1+C_{2,C}) > 0
\end{eqnarray*}

Therefore, 
\[
 D>0.
\]
